\documentclass[aps,prl,twocolumn,superscriptaddress,floatfix,showkeys,longbibliography]{revtex4-2}
\usepackage{graphicx}  % Include figure files
\usepackage{subfigure}
\usepackage{dcolumn} % Align table columns on decimal point
\usepackage{bm}% bold math
\usepackage{amssymb}   % for math
\usepackage{amsfonts}   % for math
\usepackage{urwchancal}  % for small mathcal
\DeclareMathAlphabet{\mathpzc}{OT1}{pzc}{m}{it} % for small mathcal
\usepackage{comment}
\usepackage[colorlinks,linkcolor=blue,anchorcolor=blue,citecolor=blue,urlcolor=blue]{hyperref}
\usepackage{xcolor} 
\usepackage{soul}
\usepackage{tabularx,booktabs,ragged2e,multirow}
\newcolumntype{C}{>{\centering\arraybackslash}X}
\usepackage{multirow}
\usepackage{amsmath}
\usepackage{mathrsfs}
\usepackage{mathcomp}
\usepackage{textcomp}
\usepackage{dsfont}
\usepackage{esint}
\usepackage{braket}
\usepackage{textgreek}
\usepackage{lipsum}
\usepackage{marvosym} 
\usepackage{centernot}
\usepackage{cancel}
\usepackage{dashrule}
\usepackage{tikz}
\usepackage{stackengine}
\stackMath

\DeclareMathOperator{\sgn}{sgn}

\begin{document}
\count\footins = 1000 % avoid long footnotes exceeding page bounds
%\preprint{APS/123-QED}

%\title{Low-energy signature and magnetic suppression of NH skin effects}
\title{Nonmonotonic Hall effect of Weyl semimetals under magnetic field}

\author{Xiao-Xiao Zhang}
\email{xxzhang@hust.edu.cn}
\affiliation{Wuhan National High Magnetic Field Center and School of Physics, Huazhong University of Science and Technology, Wuhan 430074, China}
%\affiliation{Institute for Quantum Science and Engineering, Huazhong University of Science and Technology, Wuhan 430074, China}
%\affiliation{Wuhan Institute of Quantum Technology, Wuhan 430206, China}
\affiliation{RIKEN Center for Emergent Matter Science (CEMS), Wako, Saitama 351-0198, Japan}

\author{Naoto Nagaosa}
\email{nagaosa@riken.jp}
%\affiliation{Department of Applied Physics, University of Tokyo, Tokyo 113-8656, Japan}
\affiliation{RIKEN Center for Emergent Matter Science (CEMS), Wako, Saitama 351-0198, Japan}
\affiliation{Fundamental Quantum Science Program, TRIP Headquarters, RIKEN, Wako, Saitama 351-0198, Japan}

%\date{\today}

\newcommand{\ba}{{\bm a}}
\newcommand{\bd}{{\bm d}}
\newcommand{\bb}{{\bm b}}
\newcommand{\bk}{{\bm k}}
\newcommand{\bmm}{{\bm m}}
\newcommand{\bn}{{\bm n}}
\newcommand{\br}{{\bm r}}
\newcommand{\bq}{{\bm q}}
\newcommand{\bp}{{\bm p}}
\newcommand{\bu}{{\bm u}}
\newcommand{\bv}{{\bm v}}
\newcommand{\bA}{{\bm A}}
\newcommand{\bB}{{\bm B}}
\newcommand{\bD}{{\bm D}}
\newcommand{\bE}{{\bm E}}
\newcommand{\bH}{{\bm H}}
\newcommand{\bJ}{{\bm J}}
\newcommand{\bK}{{\bm K}}
\newcommand{\bL}{{\bm L}}
\newcommand{\bM}{{\bm M}}
\newcommand{\bP}{{\bm P}}
\newcommand{\bR}{{\bm R}}
\newcommand{\bS}{{\bm S}}
\newcommand{\bX}{{\bm X}}
\newcommand{\brho}{{\bm \rho}}
\newcommand{\cA}{{\mathcal A}}
\newcommand{\cB}{{\mathcal B}}
\newcommand{\cC}{{\mathcal C}}
\newcommand\cD{\mathcal{D}}
\newcommand{\cE}{{\mathcal E}}
\newcommand{\cG}{{\mathcal G}}
\newcommand{\cH}{{\mathcal H}}
\newcommand{\cI}{{\mathcal I}}
\newcommand{\cK}{{\mathcal K}}
\newcommand{\cM}{{\mathcal M}}
\newcommand{\cP}{{\mathcal P}}
\newcommand{\cT}{{\mathcal T}}
\newcommand{\cV}{{\mathcal V}}
\newcommand{\cm}{{\mathpzc m}}
\newcommand{\cn}{{\mathpzc n}}
\newcommand{\balpha}{{\bm \alpha}}
\newcommand{\bbeta}{{\bm \beta}}
\newcommand{\bdelta}{{\bm \delta}}
\newcommand{\bgamma}{{\bm \gamma}}
\newcommand{\bGamma}{{\bm \Gamma}}
\newcommand{\bpi}{{\bm \pi}}
\newcommand{\bzero}{{\bm 0}}
\newcommand{\bOmega}{{\bm \Omega}}
\newcommand{\bsigma}{{\bm \sigma}}
\newcommand{\bUpsilon}{{\bm \Upsilon}}
\newcommand{\bcA}{{\bm {\mathcal A}}}
\newcommand{\bcB}{{\bm {\mathcal B}}}
\newcommand{\bcD}{{\bm {\mathcal D}}}
\newcommand{\Domega}{ {\Delta\omega}}
\newcommand{\Dbk}{ {\Delta\bk}}
\newcommand\dd{\mathrm{d}}
\newcommand\ii{\mathrm{i}}
\newcommand\ee{\mathrm{e}}
\newcommand\bark{\bar{k}}
\newcommand\bbark{\bar{\bm k}}
\newcommand\zz{\mathsf{z}}
\newcommand\colonprod{\!:\!}

\makeatletter
\let\newtitle\@title
\let\newauthor\@author
\def\ExtendSymbol#1#2#3#4#5{\ext@arrow 0099{\arrowfill@#1#2#3}{#4}{#5}}
\newcommand\LongEqual[2][]{\ExtendSymbol{=}{=}{=}{#1}{#2}}
\newcommand\LongArrow[2][]{\ExtendSymbol{-}{-}{\rightarrow}{#1}{#2}}
\newcommand{\cev}[1]{\reflectbox{\ensuremath{\vec{\reflectbox{\ensuremath{#1}}}}}}
\newcommand{\red}[1]{\textcolor{red}{#1}} %for displaying red texts
\newcommand{\blue}[1]{{\leavevmode\color{blue}#1}} %for displaying blue texts
\newcommand{\green}[1]{\textcolor{orange}{#1}} %for displaying blue texts
\newcommand{\mytitle}[1]{\textcolor{orange}{\textit{#1}}}
\newcommand{\mycomment}[1]{} %for commenting out
\newcommand{\note}[1]{ \textbf{\color{blue}#1}}
\newcommand{\warn}[1]{ \textbf{\color{red}#1}}

\makeatother

\begin{abstract}
Hall effect of topological quantum materials often reveals essential new physics and possesses potential for application. Magnetic Weyl semimetal is one especially interesting example that hosts an interplay between the spontaneous time-reversal symmetry-breaking topology and the external magnetic field. However, it is less known beyond the anomalous Hall effect thereof, which is unable to account for plenty of magnetotransport measurements. We propose a new Hall effect characteristically nonmonotonic with respect to the external field, intrinsic to the three-dimensional Weyl topology and free from chemical potential fine-tuning. Two related mechanisms from the Landau level bending and chiral Landau level shifting are found, together with their relation to Shubnikov-de Hass effect. This field-dependent Hall response, universal to thin films and bulk samples, provides a concrete physical picture for existing measurements and is promising to guide future experiments.
\end{abstract}

%\pacs{71.10.Pm, 71.27.+a, 72.15.Nj, 72.15.Rn}
%\keywords{quantum Hall effect, Weyl semimetal, magnetotransport, anisotropy, three dimensions}

\maketitle
%\tableofcontents
% \newpage
% \clearpage

%\clearpage
\let\oldaddcontentsline\addcontentsline% Store \addcontentsline
\renewcommand{\addcontentsline}[3]{}% Make \addcontentsline a no-op

%\section{Introduction}
\mytitle{Introduction}.--
Hall effect and related magnetotransport have long been a constant source of new physics for solid-state systems. The recent surge of topological quantum materials significantly invigorated %and enriched 
the relevant transport study.
Weyl semimetal (WSM) is a three-dimensional (3D) epitome and surpasses %simply extending 
the two-dimensional (2D) Dirac physics\cite{Volovik1987,Weyl2007,Weyl2011,TaAs1,TaAs2,ReviewYan2017,Burkov2018,WeylDiracReview,Nagaosa_2020,Lv2021}. 
%Its linear band crossings as Weyl points (WPs) bring about many interesting phenomena, e.g., various longitudinal chiral anomaly effects under a magnetic field\cite{Nielson-Ninomiya2,CME1,Burkov2014,seeCMEDirac1,seeCMEDirac2,seeCMEWeyl1,Hirschberger2016,Liang2018a,Cheng2021,Ong2021}. 
%Weyl orbit combining arc and bulk states\cite{Potter2014,Wang2017a,Zhang2018}.
Its linear band crossings as Weyl points (WPs) and the unique topology bring about many interesting phenomena.
Without an applied magnetic field, the surface circulating states due to slices between WPs can produce an anomalous Hall effect (AHE) as a hallmark of magnetic WSMs\cite{Ran_model,AHE2}. Such time-reversal $\cT$-breaking systems were predicted theoretically\cite{WeylwithP1,WeylwithP2,Wang2016,Yang2017,Jin2017,Nie2017,Xu2018,Borisenko2019,Nie2020} with %great interest due to 
an appealing interplay between magnetism and topology, and experimentally investigated in Mn$_3$Sn\cite{Nakatsuji2015,Li2023},
GdPtBi\cite{Hirschberger2016,Suzuki2016,Shekhar2018,Liang2018a},
%YbMnBi$_2$\cite{Borisenko2019},
Co$_3$Sn$_2$S$_2$\cite{Liu2018a,Wang2018a,Liu2019,Morali_2019,Jiang2021}, etc.

The magnetotransport thereof in the Hall channel is usually crudely decomposed as the magnetic field $B$-linear ordinary Hall effect (OHE) and the magnetization-induced AHE. This, however, often fails to account for the rich field-dependent Hall effects observed. They are typically nonlinear and nonmonotonic in $B$ and not due to hysteresis, whose mechanism remains unclear\cite{Suzuki2016,Shekhar2018,Liu2018a,Yang2020b,Li2020b,Wu2023}. Theoretically, this oversimplified picture also contradicts the intuition that the WSM topology should enter more nontrivially into the Hall effect at finite fields. 
This becomes especially true when the quantum effects of magnetic fields are properly taken into account with Landau level (LL) formation. WSMs with relativistic band structure and intriguing 3D topology should naturally accommodate special LLs, associated with nontrivial field dependence and influence on the Hall response\cite{WeylDiracReview,Burkov2018}. %In the present study, 
Here, we address this issue and propose a nonmonotonic field-dependent Hall effect in a magnetic WSM, 
combining the intrinsic $\cT$-breaking topology and the field-induced quantum effects of LLs. Given the universal mechanism, this Hall effect is present in both thin films and bulk samples; neither does it require fine-tuning the chemical potential close to WPs while many other transport features of WSM demand. It therefore represents a new Hall effect at the crossroad between topology, spontaneous $\cT$-breaking, and external magnetic field. By revealing this Hall effect in comparison to experimental results, a promising physical picture for the foregoing measurements is provided.

\begin{figure*}[!htbp]
\includegraphics[width=17.8cm]{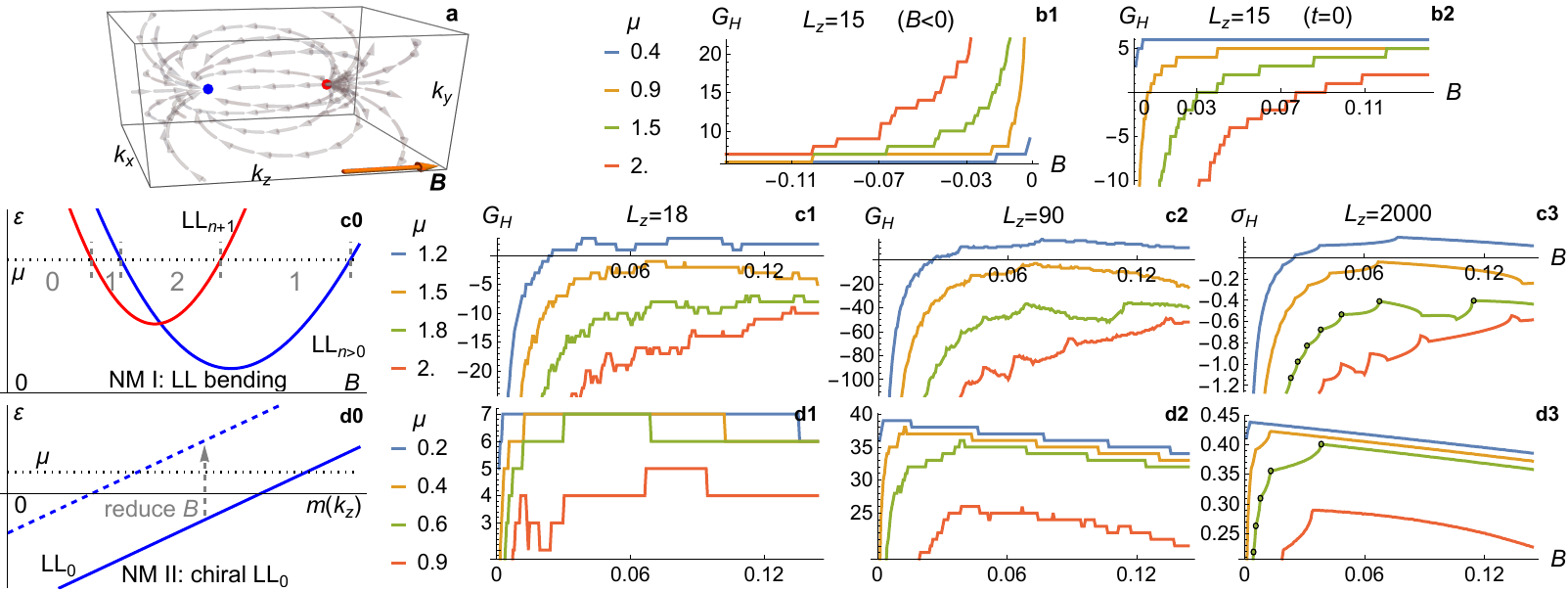}
\caption{(a) Schematic model system: a pair of positive (red) and negative (blue) WPs are aligned along $k_z$-axis; gray Berry curvature streamlines connect WPs and represent intrinsic $\cT$-breaking topology, which is at interplay with the orange external magnetic field $\bB\|\hat{z}$. 
(b) Purely monotonic Hall signals when (b1) $B<0$ or (b2) $t=0$ in contrast to their nonmonotonic counterpart (c1,d1). 
(c0,d0) Both nonmonotonic mechanisms (NM~I and II) of Hall effect originate from intrinsic WSM topology. 
(c0) Represented by two bent and reshuffled LL$_n$ and LL$_{n+1}$, NM~I exploits the anomalously nonmonotonic occupation change (numbers below dotted chemical potential $\mu$) when $B$ is reduced; (d0) NM~II relies on the field-induced shift of chiral LL$_0$ and the associated occupation change in relevant $m(k_z)$-slices. 
(c1-3,d1-3) Nonmonotonic Hall signals dependent on chemical potential $\mu$ and thickness $L_z$: (c1-2,d1-2) thin films or mesoscopic samples (conductance $G_H$ in units of $e^2/\hbar$); (c3,d3) bulk samples (3D conductivity $\sigma_H=G_H/L_z$).
Parameters are $\chi=-1,t_1=3,t_0=4,t=12,k_w=\pi/2$. 
(c1-3) Higher $\mu$ clearly shows NM~I often with irregular nonmonotonicity while NM~II also participates. 
(d1-3) As $\mu$ decreases across the minimal threshold energy $\cM_1=0.75$ for NM~I, generally more irregular %$\mu>\cM_1$ 
signals transition to regular NM~II signals with one major nonmonotonic turn.  
(c3,d3) clearly show the Shubnikov-de Haas effect (SdHE) with successive cusps, circled representatively in green curves, modified by nonmonotonicity to acquire unconventional periodicity.
}\label{Fig:main}
\end{figure*}

%\section{Model system and Hall response}
\mytitle{Model system and Hall response}.--
We start from a minimal $\mathcal{T}$-breaking WSM lattice Hamiltonian $H(\bk)=t_1(\chi\sin{k_x}\sigma_x+\sin{k_y}\sigma_y) + [t_0(\cos{k_z}-\cos{\bar k_z})+t\sum_{i=x,y}(\cos{k_i}-\cos{\bar k_i})]\sigma_z$
with chirality $\chi=\pm$, Pauli matrix $\sigma$ and a pair of zero-energy WPs at $\pm\bar\bk=(0,0,\pm k_w)$ illustrated in Fig.~\ref{Fig:main}(a). For concreteness, we assume $t_0,t_1,t>0$ and leave other cases to the Supplemental Material (SM). The corresponding continuum model we mainly rely on reads
\begin{equation}\label{eq:h_main}
    h(\bk)=t_1(\chi k_x\sigma_x+k_y\sigma_y) + [m(k_z)-t(k_x^2+k_y^2)/2]\sigma_z
\end{equation}
with $m(k_z)=m_0-\frac{t_0}{2}k_z^2$ and $m_0=t_0(1-\cos{k_w})$.
Eq.~\eqref{eq:h_main} differs from usual Dirac models by including $t$, the natural high-energy regularization from the lattice. 
Only by this the essential lattice WSM topology is properly captured, otherwise it would suffer from parity anomaly and half-quantized Chern number for any 2D $k_z$-section\cite{Niemi1983,*Redlich1984,Mogi2022}.
With $t$, Chern numbers are integer-quantized: topologically nontrivial (trivial) between (outside) the WPs.
This complete global topology of realistic WSMs will significantly impact Hall effects, later proved to be an indispensable ingredient of any nonmonotonic behavior. 
Applying a magnetic field $B\hat{z}$ for a sample of thickness $L_z$ along $z$-axis, 
the allowed discrete $k_z$'s are conserved and Eq.~\eqref{eq:h_main} becomes an effective 2D model of each $m(k_z)$-slices. 
The general LL$_{n\neq0}$ reads $\varepsilon_{n}=\bar\varepsilon+\sgn(n)\sqrt{(m-2|n| u_{B})^2+|n|v_{B} ^2}$ with $m=m(k_z)$ while the special LL$_0$ $\varepsilon_{0}=\bar\varepsilon-\chi bm$ is chiral in the $(m,\varepsilon)$-plane, where $\bar\varepsilon=\chi b u_{B}, u_{B}=t/2l_{B}^2,v_{B} =\sqrt{2t_1^2/l_{B}^2}$ and magnetic length $l_{B}=\sqrt{\hbar/e|B|}$ with 
$b=\sgn(B)$ the sign of the field.
%$b=\sgn(B),|B|=|B|$ the sign and magnitude of the field.

Detailed in \ref{App:Hall}, the Hall conductance of Eq.~\eqref{eq:h_main} for a particular $m(k_z)$-slice can be separated as $G_H=\bar{G}_H+\Delta G_H$. Measured in units of $e^2/h$,  %$\bar{G}_H=\mycomment{\frac{e^2}{h}}\frac{-\chi}{2}[\sgn(m)+\sgn(t)]$ 
\begin{equation}\label{eq:barsigmaHchi_main}
    \bar{G}_H= \mycomment{\frac{e^2}{h}}-\chi[\sgn(m)+\sgn(t)]/2,
\end{equation}
independent of the chemical potential $\mu$ and $B$ and due to the intrinsic topology, is apparently the same as the zero-field Hall effect from topological insulator thin films and contributes to the foregoing AHE in WSMs\cite{Lu2010}. 
\begin{equation}\label{eq:DeltasigmaH_main}
    \Delta G_H=\mycomment{\frac{e^2}{h}}-b\nu \Theta(\Delta)
        \Theta(n_L-n_R) \,(n_L-n_R+1)
\end{equation}
reflects the topology influenced by magnetic field and occupation, where $n_L=\max{[-1,\lfloor n_+\rfloor]},n_R=\max{[\xi ,\lceil n_-\rceil]}$ and $n_\pm=\frac{4m u_{B}-v_{B} ^2\pm\sqrt{\Delta}}{8 u_{B}^2}$
with $\Delta=v_{B} ^2(v_{B} ^2-8m u_{B})+16 u_{B}^2(\mu-\bar\varepsilon)^2$, $\xi =\frac{1+\sgn(m)\chi b\nu }{2}\mycomment{[1+\sgn(m)\chi b\nu ]/2}$, and $\nu =\sgn(\mu-\bar\varepsilon)$ switching sign between electron-like and hole-like contributions. 
$\Theta(x)$ is the step function and $\lfloor x \rfloor$ ($\lceil x \rceil$) denotes the greatest (smallest) integer less (greater) than or equal to $x$. 
$n_{L,R}$ signifies the proper LL index regarding the left/right crossing of chemical potential. The physical motivation is to count LLs filled with respect to the gapped region near band midpoint; $\Theta$-functions guarantee such fillings and associated quantized contributions from LLs are picked up.
With these analytic formulae, one can sum over the $m(k_z)$-slices and find the full conductance $G_H=\sum_{k_z} G_H(m(k_z))$ dependent on $L_z$ measured in lattice constant, as shown in Fig.~\ref{Fig:main}. %Expected from realistic systems, there always exists a lower bound $m^*<0$ of $m(k_z)$ such that $ G_H(m<m^*)\equiv0$ (\ref{App:3Dformula}), which naturally stops the momentum summation and, interestingly, implies that Hall responses are contributed even outside the WPs ($m=0$), in contrast to the AHE.
$G_H$ satisfies
\begin{equation}\label{eq:symm_main}
    G_H(B,\mu)=G_H(-B,-\mu)
\end{equation}
related to the particle-hole symmetry breaking and the Onsager reciprocal relations (\ref{App:symmetry}). 
The $B$-asymmetry $G_H(B,\mu)\neq G_H(-B,\mu)$ implied in Eq.~\eqref{eq:symm_main} originates from the $\cT$-breaking in our magnetic WSM and will manifest drastically in the nonmonotonic Hall effect. 

In the conventional 2D electron gas or graphene\cite{Review:graphene}, %LL energies monotonically move with $|B|$. 
the number of filled/empty LLs always increases with decreasing $|B|$ as they are pushed toward the band midpoint with monotonically decreasing LL spacings. 
This leads to a common monotonic $G_H(B)$ as in the quantum Hall effect.
When $b\mu<0$ or $t=0$, Figs.~\ref{Fig:main}(b1,b2) follow the same conventional behavior (see also SM Fig.~\ref{Fig:other}). Remarkably, when $b\mu>0$, Figs.~\ref{Fig:main}(c,d) stably exhibit a nonmonotonic $G_H(B)$ in a wide field window, except the downturn at small $B$ also due to the conventional LL spacing effect. Note, however, that such apparent divergence with $B
\rightarrow0$ will only be regularized to the OHE as $G_H
\propto\omega_c\tau\mycomment{\frac{e^2}{\hbar}}\propto B$ with cyclotron frequency $\omega_c$ and relaxation time $\tau$ due to realistic scatterings (\ref{App:LLspacing}).

%\section{Nonmonotonic mechanism I: LL bending and reshuffling}%\label{App:LLreshuff}
\mytitle{Nonmonotonic mechanism I: LL bending and reshuffling}.--
To understand the nonmonotonic behavior, one notices a unique LL$_{n\neq0}$ bending phenomenon of WSM exemplified in Fig.~\ref{Fig:main}(c0), i.e., sign switching of LL slope %$\frac{\partial\varepsilon_n}{\partial|B|}$ 
or nonmonotonic movement of LL with $|B|$.
Once any filled LL bends upward with decreasing $|B|$, it is possible to be emptied when the fixed $\mu$ crosses it again at a lower $|B|$, resulting in the opposite Hall contribution. 
With this nonmonotonic mechanism (denoted as NM~I), there are multiple nonmonotonic turns, dependent intricately on $B,\mu$, the variable momentum slice $m(k_z)$, and the participating LLs. Hence the robust but often irregular nonmonotonic behavior in Figs.~\ref{Fig:main}(c1-3). 
We find a sufficient condition for NM~I (\ref{App:LLbending_cond})
\begin{equation}\label{eq:m&mu_condition}
    %m_0>
    m(k_z)>-\chi b\mu > \cM_n
\end{equation}
with a threshold energy $\cM_n\mycomment{=|n|\cm_0}=|n|t_1^2/t$ linked to LL index $n$. %A common mass maximum $m_0$ exists per the Hamiltonian. 
Since $\cM_n$ increases with $|n|$, any LL$_{|l|\leq|n|}$ can participate NM I, as summarized in Table~\ref{tab:NM_SdH_relation}.
%As physically expected, the constraint of $\mu$ reflects that neither too high nor too low $|\mu|$ can pick up the desired intricate filling change: $m$, the higher bound of $|\mu|$, is due to the singularity at $B=0$ that restricts the bending of LLs; $\cM_n$, the lower bound of $|\mu|$, asks the chemical potential to at least cross LL$_n$. 
Intuitively, to pick up the nonmonotonic occupation change in Fig.~\ref{Fig:main}(c0), $\mu$ must cross LLs; the constraint on $|\mu|$ from $\cM_n$ to $m(k_z)$ follows from the relevant LL extrema.
%Its lower bound $\cM_n$ is related to LL extremum; its upper bound $m(k_z)$ is because $\mu$ cannot cross LL beyond the ending at $B=0$. %$\mu$ crosses LL$_n$ while the crossings should not exceed the physical range since bending LLs always end at $B=0$. 
Bending LLs are often accompanied by the reshuffling of LL ordering, e.g., anomalous $\varepsilon_{n+1}<\varepsilon_n$ exemplified in Fig.~\ref{Fig:main}(c0). This complementary view can retain explicit $|B|$-dependence and is explored in \ref{App:LLslope}.

\newcommand{\blap}[1]{\vbox to 13.5pt{\hbox{#1}\vss}}

\begin{table}[!hbtp]
    \centering
    \begin{tabular}{c|c|c|c}
 & \blap{\shortstack{NM~I\\(LL bending)}} & \blap{\shortstack{NM~II\\(chiral LL$_0$)}} & SdHE \\
 \cmidrule{1-4}
         $-\chi b\mu\leq0$ &  - & - & all LL: \checkmark \\
         $0<-\chi b\mu<\cM_n$ &  LL$_{|l|\geq|n|}$:  - &  \checkmark & LL$_{|l|\geq|n|}$: \checkmark \\
         $-\chi b\mu>\cM_n$ &  LL$_{|l|\leq|n|}$: \checkmark & \checkmark & LL$_{|l|\leq|n|}$: -\\
    \end{tabular}
    \caption{Relations among two nonmonotonic mechanisms and SdHE. They summarize Eqs.~\eqref{eq:m&mu_condition}, \eqref{eq:osci_cusp_exist}, and \eqref{eq:NMLL_0_condition}. 
    Leftmost conditions are specified with a threshold energy $\cM_n$ linked to LL index $n$ and dependent on chemical potential $\mu$, WSM chirality $\chi$, and sign of magnetic field $b$. 
    Fixing such conditions, presence/absence (\checkmark/-) of corresponding phenomena due to generic LL$_l$ can be given in relation to fixed $n$.
    NM~I and SdHE due to the same LL (distinct LLs) cannot (can) coexist.
    NM~II persists regardless of $\cM_n$.
    Nonmonotonic Hall effect disappears when $-\chi b\mu\leq0$.}
    \label{tab:NM_SdH_relation}
\end{table}

Crucially, it is the quadratic $t$-term in Eq.~\eqref{eq:h_main} that enables LL bending/reshuffling in Fig.~\ref{Fig:main}(c0). Without it, LLs do not bend, divergent $\cM_n$ renders the condition Eq.~\eqref{eq:m&mu_condition} never fulfilled, and transport returns to the conventional purely monotonic behavior shown in Fig.~\ref{Fig:main}(b2). As aforementioned, finite $t$ is essential to capture the nontrivial and realistic global WSM topology. %also captured in the AHE-like Eq.~\eqref{eq:barsigmaHchi_main}.
Therefore, NM~I is a direct and intrinsic signature of the WSM topology under magnetic field.

Eq.~\eqref{eq:m&mu_condition} immediately suggests more bending LLs accommodated and hence typically stronger nonmonotonicity with higher $|\mu|$, corroborated by Figs.~\ref{Fig:main}(c1-3) and also its comparison with (d1-3). Besides, reducing $\mu$ eventually across the threshold minimum $\cM_1$, Figs.~\ref{Fig:main}(d1-3) shows a pattern shift to a distinctly more regular nonmonotonicity (discussed later).
Eq.~\eqref{eq:m&mu_condition} also manifests the aforementioned $B$-asymmetry %between $b>0$ and $b<0$ 
in a remarkable way. 
%\textit{Only} the symmetry Eq.~\eqref{eq:symm_main} related $b>0,\mu>0$ and $b<0,\mu<0$, e.g., when $\chi=-1$ as exemplified, can significantly exhibit nonmonotonicity. 
Exemplifying with $\chi = -1$, only $b>0,\mu>0$ and $b<0,\mu<0$ related by Eq.~\eqref{eq:symm_main} can show the nonmonotonicity while other cases rarely do  (\ref{App:exceptions}). 
This thus becomes a drastic feature of the $\mathcal{T}$-breaking in the WSM system. 

%Note that $\mu$ has to be roughly comparable to the reordered LLs, e.g., their crossing energy, at the given $m(k_z)$ in order to take advantage of the reshuffling; .

% %As aforementioned, the $t=0$ case resembles the conventional scenario without LL bending. 
% Finite $t$, responsible for the global WSM topology, now crucially makes %the conditions 
% Eq.~\eqref{eq:m&mu_condition} %\eqref{eq:reshuffle_condition_main} 
% possible. Without $t$, divergent $\cM_n$ render the condition never fulfilled: it returns to the conventional scenario without LL bending/reshuffling. %On the other hand, finite $t$ in Eq.~\eqref{eq:h_main} is directly related to the WSM topology and its $\mathcal{T}$-breaking, i.e., topologically nontrivial $k_z$ slices only exist between the pair of WPs, which is as well captured by Eq.~\eqref{eq:barsigmaH} or Eq.~\eqref{eq:barsigmaH_B=0}. 
% As aforementioned, finite $t$ in Eq.~\eqref{eq:h_main} is essential to the nontrivial and realistic WSM topology. %also captured in the AHE-like Eq.~\eqref{eq:barsigmaHchi_main}.
% Therefore, NM~I is a direct and intrinsic signature of the WSM topology under magnetic field.

%\section{Shubnikov-de Haas effect}
\mytitle{Shubnikov-de Haas effect}.--
During the reduction of $|B|$, the Hall signal especially for a bulk sample will also exhibit a Shubnikov-de Haas effect (SdHE). Such contribution from LL spacing reduction is itself monotonic as aforementioned and thus causes cusps rather than direct oscillations as shown in Figs.~\ref{Fig:main}(c3,d3). A cusp at $|B|_{n}$ %$|B|_{n}\mycomment{=\frac{2\hbar}{e}\frac{2|n|t_1^2+t\chi b\mu - 2t_1\sqrt{|n|(|n|t_1^2+t\chi b\mu)}}{t^2}}=\frac{2\hbar}{et}[\cM_{2n}+\chi b\mu - 2\sqrt{\cM_n(\cM_n+\chi b\mu)}]$ 
appears once $\mu$ touches the extremum of a particular LL$_{n\neq0}$ in any $m(k_z)$ slice, provided that (\ref{App:SdH_cusp}) 
\begin{equation}\label{eq:osci_cusp_exist}
    %|n|t_1^2+tb\mu>0
    -\chi b\mu<\cM_n.
\end{equation}
The foregoing finite $t$ crucial to WSM makes the cusp separation $\Delta |B|_n^{-1}$ nonuniform, in contrast to the conventional case.

Intriguingly, Eq.~\eqref{eq:osci_cusp_exist} and Eq.~\eqref{eq:m&mu_condition} from two distinct viewpoints are mutually exclusive. Their connection can be rigorously built by considering %the chemical potential $\mu$-cut of 
the LL energy surface in the 3D $(m,|B|,\varepsilon)$-space and its extremal curves (\ref{App:SdHEandLLbending}). 
Intuitively speaking, the LL bending effect requires $\mu$ crossing an LL and hence it hardly reaches the extremum to induce any SdHE kinks.
This important relation enriches Table~\ref{tab:NM_SdH_relation} for two nonmonotonic mechanisms and SdHE.
It is, however, worth pointing out that the two \textit{phenomena}, SdHE and nonmonotonicity due to LL bending, are not exclusive. %Eq.~\eqref{eq:osci_cusp_exist} means that 
SdHE cusps due to LL$_{|l|\geq|n|}$ can occur while LL$_{|l|<|n|}$ bending may still coexist if parameters allow $-\chi b\mu>\cM_{|l|<|n|}$, e.g., one can observe SdHE following nonmonotonicity as $|B|$ decreases, seen in Fig.~\ref{Fig:main}(c3). This clarifies the no-go $(2,1)$-entry and the similar $(3,3)$-entry in Table~\ref{tab:NM_SdH_relation}.

%\section{Nonmonotonic mechanism II: Chiral \texorpdfstring{LL$_0$}{LL0}}\label{App:LL0}
\mytitle{Nonmonotonic mechanism II: Chiral LL$_0$}.--
As aforementioned and clearly suggested by Figs.~\ref{Fig:main}(d1-3), there exists another indispensable nonmonotonic mechanism (NM~II). It gives rise to a distinctly more regular nonmonotonic variation in $G_H(B)$ than NM~I: the profile typically exhibits one overall nonmonotonic turn before approaching the $|B|\sim0$ region, although it possibly consists of multiple steps for a thin film as in Figs.~\ref{Fig:main}(d1,d2). Illustrated in Fig.~\ref{Fig:main}(d0), this originates from the unique chiral LL$_0$ movement  %seen from Eq.~\eqref{eq:DepsilonDB}, 
that shifts linearly with $|B|$ as per $\frac{\partial\varepsilon_0}{\partial|B|}=e t\chi b/2\hbar$. Given $\mu$, this changes the occupation in momentum slices and hence $G_H$ in a nontrivial way (\ref{App:LLspacing}):
%upward/downward linearly as a whole with decreasing $|B|$. %respectively for $\chi b=\mp$. 
%Correspondingly for a given quantized $m(k_z)$ slice and decreasing $|B|$, $\Delta G_H(B,\mu,m(k_z))$ will change by $-\chi\frac{e^2}{h}$ once LL$_0$ moves across $\mu$ %, regardless of $b$ (\ref{App:LLspacing}). 
the change %upon $G_H$ 
is opposite to the foregoing conventional effect of LL spacing reduction %when $\chi b\sgn(\mu)=-$.
%This effect of decreasing $G_H$ is constructive (destructive) to the \red{foregoing} conventional shrinking LL spacing effect of decreasing (increasing) $G_H$ respectively when $b\sgn(\mu)=\pm$. [Or when $\chi=-1$ this effect of increasing $G_H$ is destructive (constructive) to the foregoing conventional shrinking LL spacing effect of decreasing (increasing) $G_H$ respectively when $b\sgn(\mu)=\pm$.] 
if and only if
\begin{equation}\label{eq:NMLL_0_condition}
    -\chi b\mu>0.
\end{equation}
These two opposite contributions to $G_H(B)$ naturally lead to nonmonotonicity and its condition above complies with Eq.~\eqref{eq:symm_main}. 
NM~II will also modify and compete with SdHE, because SdHE %, due to the conventional LL spacing variation, 
follows the monotonic behavior. 
%Their competition not only contributes to the nonmonotonic signal, but also can shift leftward the cusps $|B|_n$ predicted by pure SdHE effects, e.g., in Fig.~\ref{Fig:main}(d3).

Eq.~\eqref{eq:NMLL_0_condition} is also a necessary condition of Eq.~\eqref{eq:m&mu_condition} for LL bending.
Hence, NM~I and II are inclusive of each other under Eq.~\eqref{eq:NMLL_0_condition} and will contribute constructively together when the system parameters permit as in Figs.~\ref{Fig:main}(c1-3). 
NM II is not limited by the mass range $m_0>\cM_1$ implicitly implied by Eq.~\eqref{eq:m&mu_condition} and can always contribute. %as long as Eq.~\eqref{eq:NMLL_0_condition} holds.
Especially, if $|\mu|$ or $t$ is too small to fulfill Eq.~\eqref{eq:m&mu_condition}, 
typically when $|\mu|$ is close to or below the threshold minimum $\cM_1$, 
NM~II becomes the major cause of nonmonotonicity, as aforementioned and shown in Figs.~\ref{Fig:main}(d1-3). 

It is important to realize that NM~II is also a direct consequence of the same finite $t$ as NM~I. It now generates the field-dependent shift of LL$_0$ in Fig.~\ref{Fig:main}(d0) and is crucial to nonmonotonicity as suggested by Fig.~\ref{Fig:main}(b2), where neither of two NMs exist. Again, the nonmonotonicity traces intrinsically back to the generic 3D global WSM topology. In fact, although both NMs considerably rely on 2D Dirac physics, WSM provides the natural embedding in 3D with the requisite regularization and a whole family of different 2D slices. This crucially removes any necessity of fine-tuning and solidifies nonmonotonicity, in a way reminiscent of the generic appearance of WPs in 3D.

%\section{Thick sample situation}
\mytitle{Thick sample situation}.--
For a bulk sample of large $L_z$, the allowed quantized $k_z$'s form a continuum and we have instead the bulk 3D conductivity $\sigma_H=G_H/L_z=\frac{1}{\pi}\int\dd k_z  G_H(m(k_z))$. As clearly shown in Figs.~\ref{Fig:main}(c3,d3), NM~I and II are not limited to a thin sample of a few atomic layers or mesoscopic scale. Traversing the relevant LLs, the integral will stably pick up both the conventional monotonic part and the two nonmonotonic contributions.
However, the conductance variation in discrete steps, which strongly depend on the system details and thickness, will be considerably smoothed out. 
Therefore, not only do NM~I and II survive, but the Hall signal can become simpler with fewer irregular nonmonotonic changes.

There are also differences between NM~I and II. NM~II is scale-invariant due to the chiral feature of LL$_0$, making NM~II almost intact from parameter changes. Only the signal becomes less discrete and smoother in thicker samples as in Figs.~\ref{Fig:main}(d1-3). On the other hand, NM~I can be associated with a typical field strength $\cB_0=\hbar/el_0^2$ with $l_0=2t/t_1$. %, which approximately corresponds to the change in magnetic field for LL bending and is better not large compared to the maximal applied field. 
Shown in Figs.~\ref{Fig:main}(c1-3), although a mesoscopic thickness well hosts both NM~I and II, NM~I in thick samples may become less outstanding for small fields %$l_0\ll l_{B}$, 
$|B|\ll\cB_0$. 
In such a case, the monotonic contribution is densely accumulated through continuous momentum slices, which may outrun the contingent nontrivial occupation changes in Fig.~\ref{Fig:main}(c0). 
However, as per Eq.~\eqref{eq:m&mu_condition}, higher $|\mu|$ usually recovers and enhances nonmonotonicity by involving more bending LLs, e.g., in Fig.~\ref{Fig:main}(c3).
Therefore, while NM~II remains exceptionally stable, a higher field or %hopping $t$ and 
$|\mu|$ is beneficial to enhancing NM~I in thicker samples.

%\section{Implications on experiments}
\mytitle{Implications on experiments}.--
Following such discussion, the magnetic field observation window for significant NM~I is estimated to be approximately $10\text{-} 60\mathrm{T}$ with the typical $\cB_0\sim30\mathrm{T}$\cite{Nie2020,Muechler2020,Wang2012a} (\ref{App:Parameters}). This is reachable with laboratory magnets and especially pulsed magnetic fields, which routinely go beyond $50\mathrm{T}$ and is widely used in magnetotransport\cite{Fauque2013,Uchida2021,Kondo2023,Wu2023}. On the other hand, NM~II persists all along and will dominate more for $\mu$ closer to WPs. 
Besides, the nonmonotonic Hall effect is limited by $|B|\gtrsim|B|_c\sim\hbar^2m_0/e\tau t_1^2$. When $\omega_c\tau<1$ and relaxation becomes important, eventually the OHE prevails over quantum effects. Together with the realistic nondivergence that $G_H(B\rightarrow0)$ approaches the AHE, they crucially regularize and suppress the apparently divergent downturn at small fields in Figs.~\ref{Fig:main}(c,d) (\ref{App:LLspacing}).  
This naturally benefits the nonmonotonic effect to reach smaller fields beyond the present calculation, which mainly accounts for the $\omega_c\tau>1$ clean case.

Given that $\mu$ can vary in different materials or due to chemical doping, one might also observe the nonmonotonic $b<0,\mu<0$ case complementary to Fig.~\ref{Fig:main}(c,d). 
In typical magnetic WSMs, the intrinsic magnetization withstands an antiparallel magnetic field up to a coercive field $H_c$ about $0.1\textrm{-}10\mathrm{T}$\cite{Nagaosa_2020,Liu2018a,Li2020b,Thakur2020,Ikeda2021,Yoshikawa2022}. If the onset field of LL is larger, one cannot enter deep into the nominal $b<0$ regime for our system 
exhibiting %an overall 
magnetization in $\hat{z}$. However, once the magnetization is reversed 
by negative $B$, the inequalities Eqs.~\eqref{eq:m&mu_condition}\eqref{eq:NMLL_0_condition} will also be \textit{reversed} and nonmonotonicity is identically allowed (\ref{App:LLbending_cond}). This relaxes the constraint of $B,\mu$ in experiments. 

As aforementioned, various recent Hall measurements in magnetic WSMs show anomalously nonmonotonic (sometimes termed `nonlinear') field dependence, beyond the OHE and AHE. They are typically observed in the regime $\omega_c\tau$ larger than or close to unity, where quantum effects appear and the major contribution is found not from extrinsic scatterings\cite{Suzuki2016,Liu2018a,Shekhar2018,Wu2023}. There has been no clear and consistent understanding, except, e.g., phenomenological fittings with electron and hole carriers. Our prediction qualitatively matches them and enables a promising physical picture: GdPtBi\cite{Suzuki2016,Shekhar2018} and Co$_3$Sn$_2$S$_2$ with magnetization reversal measurements\cite{Liu2018a,Yang2020b,Li2020b} typically confirm one nonmonotonic turn similar to Fig.~\ref{Fig:main}(d3) of NM~II when $|B|<10\text{-}15\mathrm{T}$; PrAlSi at pulsed fields higher than $15\mathrm{T}$\cite{Wu2023} exhibits more complex nonmonotonic variation as Fig.~\ref{Fig:main}(c3) of NM~I suggests, which corroborates our estimation; PrAlSi\cite{Wu2023} and GdPtBi\cite{Suzuki2016} suggest deviation from conventional SdHE. Experiments also confirm the nondivergence at small $|B|$ as we envisaged. \ref{App:existing_experiment} presents a more thorough comparison and analysis. Despite the resemblance between these experiments and the theory, one should still keep in mind other possible contributions, e.g., more complex variation of AHE induced by magnetic field\cite{Nagaosa2010,Takahashi2018}. 

In those systems with more than one pair of WPs, we can essentially consider multiple copies of the present model. Similar nonmonotonic contributions from other pairs will %appear due to the same mechanisms and 
add up together in Hall signals. Given field $\bB$, it is those momentum slices perpendicular to $\bB$ 
that contribute, as long as $\bB$ is not orthogonal to the WP pair alignment. It is, however, intriguing to note that Eq.~\eqref{eq:NMLL_0_condition} can render some WP pairs ineffective. This happens, e.g., when we have two pairs with opposite chirality-$\chi$ arrangements, one is thus not functioning. In other words, by observing nonmonotonicity experimentally, this chirality-selectiveness can also help determine the topological configuration of WP pairs (\ref{App:selective}).

We lastly emphasize an advantage of the present phenomenon. Unlike many other transport features of WSM that typically rely on $\mu$ pinned to  WPs, fine-tuning of $\mu$ is unnecessary. This strongly relaxes constraints on material choice and experimental setting. This remarkable property traces back to the following: the phenomenon utilizes the global topology in the WSM momentum space rather than any local feature close to WPs. For instance, with a thin film, WPs may often be gapped by the quantum confinement, against which the phenomenon remains robust.

\begin{acknowledgments}
We thank R. Yoshimi, I. Belopolski, M. Hirschberger, and Y. Xie for helpful discussions. 
This work was partially supported by the startup funding of Huazhong University of Science and Technology and JSPS KAKENHI (Nos.~24H00197 and 24H02231). N.N. was also supported by the RIKEN TRIP initiative.
\end{acknowledgments}\mycomment{\Yinyang}

%\nocite{*}

\bibliography{reference.bib}  % The references (bibliography) information are stored in the file named "Bibliography.bib"
\let\addcontentsline\oldaddcontentsline% Restore \addcontentsline

\newpage
\onecolumngrid
\newpage
{
	\center \bf \large 
	Supplemental Material \\
	%\large for ``Non-Hermitian exceptional Landau quantization in electric circuits"\vspace*{0.1cm}\\ 
	\large for ``\newtitle"\vspace*{0.1cm}\\ 
	\vspace*{0.5cm}
	%\newauthor
}
\begin{center}
    %\getauthor \\
	Xiao-Xiao Zhang$^{1,2}$ and Naoto Nagaosa$^{2,3}$\\
	\vspace*{0.15cm}
    \small{$^1$\textit{Wuhan National High Magnetic Field Center and School of Physics, Huazhong University of Science and Technology, Wuhan 430074, China}}\\
	\small{$^2$\textit{RIKEN Center for Emergent Matter Science (CEMS), Wako, Saitama 351-0198, Japan}}\\
    \small{$^3$\textit{Fundamental Quantum Science Program, TRIP Headquarters, RIKEN, Wako, Saitama 351-0198, Japan}}
	%\small{$^2$\textit{Department of Applied Physics, University of Tokyo, Tokyo 113-8656, Japan}}\\
	\vspace*{0.25cm}	
\end{center}

%\twocolumngrid	

\tableofcontents

% %\clearpage
% %\appendix
% \setcounter{equation}{0}
% \setcounter{figure}{0}
% \setcounter{table}{0}
% \setcounter{page}{1}
% %\renewcommand{\theequation}{S\arabic{equation}}
% \renewcommand{\thefigure}{S\arabic{figure}}
% \renewcommand{\bibnumfmt}[1]{[S#1]}
% %\renewcommand{\citenumfont}[1]{S#1}

%%%%%%%%%% Merge with supplemental materials %%%%%%%%%%
%%%%%%%%% Prefix a "S" to all equations, figures, tables and reset the counter %%%%%%%%%%
%\appendix
\setcounter{section}{0}
\setcounter{equation}{0}
\setcounter{figure}{0}
\setcounter{table}{0}
\setcounter{page}{1}
%\makeatletter
\renewcommand{\theequation}{S\arabic{equation}}
\renewcommand{\thefigure}{S\arabic{figure}}
\renewcommand{\thetable}{S\arabic{table}}
\renewcommand{\theHtable}{Supplement.\thetable}
\renewcommand{\theHfigure}{Supplement.\thefigure}
\renewcommand{\thesection}{SM~\Roman{section}}
\renewcommand{\bibnumfmt}[1]{[S#1]}
\renewcommand{\citenumfont}[1]{S#1}
%%%%%%%%% Prefix a "S" to all equations, figures, tables and reset the counter %%%%%%%%%%

\section{2D and 3D Hall conductances}\label{App:Hall}
\subsection{Model Hamiltonian and LLs}\label{App:HandLL} 
We consider a minimal $\mathcal{T}$-breaking WSM with a pair of WPs at $(0,0,\pm k_w)$, for which the lattice Hamiltonian reads
\begin{equation}\label{eq:H}
    H(\bk)=t_1(\chi\sin{k_x}\sigma_x+\sin{k_y}\sigma_y) + [t_0(\cos{k_z}-\cos{\bar k_z})+t\sum_{i=x,y}(\cos{k_i}-\cos{\bar k_i})]\sigma_z
\end{equation}
with $\bar\bk=(0,0,k_w)$. From this, one can further write down a continuum model 
\begin{equation}\label{eq:h}
    h(\bk)=t_1(\chi k_x\sigma_x+k_y\sigma_y) + [m(k_z)-\frac{t}{2}(k_x^2+k_y^2)]\sigma_z
\end{equation}
with $m(k_z)=m_0-\frac{t_0}{2}k_z^2$ and $m_0=t_0(1-\cos{k_w})$. For concreteness, we assume $t_0,t_1,t$ to be positive.
Note that in Eq.~\eqref{eq:H} parameters $t_1,t,t_0$ have the dimension of energy; instead, in Eq.~\eqref{eq:h} $m_0,t_1,t,t_0$ are actually $m_0,t_1a,ta^2,t_0a^2$ with lattice constant $a$ absorbed in the definition and hence change their dimensions.
Since we will apply a magnetic field $B\hat{z}$ normal to a thin film grown along $z$-axis, the allowed discrete $k_z$ values are still good quantum numbers and it is convenient to consider Eq.~\eqref{eq:h} as an effective 2D model of each $k_z$-slices.

With the electromagnetic Peierls substitution $\hbar\bk\rightarrow \bpi=-\ii\hbar\nabla+e\bA$ implemented in Eq.~\eqref{eq:h}, which satisfies $\nabla\times\bA=B\hat{z}$ with $e>0$ the elementary charge, we have a magnetic field-dependent Hamiltonian $h(B)$ and the commutation relation $[\pi_x,\pi_y]=-\ii e\hbar B$. Note that there is no need to specify a gauge here. We define the ladder operators 
\begin{equation}
c^\pm=\frac{t_1}{\hbar v_{B} }(\pi_x\pm \ii \,b\,\pi_y)
\end{equation}
with $b=\sgn(B),v_{B} =\sqrt{2 e|B| t_1^2/\hbar}=\sqrt{2t_1^2/l_{B}^2}$ and magnetic length $l_{B}=\sqrt{\hbar/e|B|}$, which satisfies $[c^-,c^+]=1$. With this, one can build the tower of states $\{\phi_{n\geq0}\}$ from $\phi_0$, which is defined by $c^-\phi_0=0$, using the relation $c^-\phi_n=\sqrt{n}\phi_{n-1},c^+\phi_n=\sqrt{n+1}\phi_{n+1}$. Our $h(B)$ is now cast into
\begin{equation}\label{eq:h^b}
h_{\chi,b}=\begin{pmatrix}
    m- u_{B}(2c^+ c^-+1) & \chi v_{B} c^{-\chi b} \\
    \chi v_{B} c^{\chi b} & -[m- u_{B}(2c^+ c^-+1)]
\end{pmatrix}
\end{equation}
% \begin{equation}\label{eq:h^b}
% h^b=\begin{pmatrix}
%     m- u_{B}(2c^+ c^-+1) & v_{B} c^-/v_{B} c^+ \\
%     v_{B} c^+/v_{B} c^- & -[m- u_{B}(2c^+ c^-+1)]
% \end{pmatrix}
% \end{equation}
with $ u_{B}=tv_{B} ^2/4t_1^2=e|B| t/2\hbar=t/2l_{B}^2$.
With the eigenstate ansatz $\psi_{ns}^{\chi b=+}=(\alpha_{ns}\phi_{n-1},\beta_{ns}\phi_n)^\mathrm{T}$ and $\psi_{ns}^{\chi b=-}=(\alpha_{ns}\phi_{n},\beta_{ns}\phi_{n-1})^\mathrm{T}$ for $s=\pm$, we find the LLs 
\begin{equation}\label{eq:LLs}
    \varepsilon_{n\pm}^{\chi b}=
    \begin{cases}
    \bar\varepsilon\pm\kappa_n & n\geq1 \\
    \bar\varepsilon-\chi bm & n=0 
    \end{cases}
\end{equation}
with $\bar\varepsilon=\chi b u_{B},\kappa_n=\sqrt{p_n^2+q_n^2},p_n=m-2n u_{B},q_n=\chi\sqrt{n}v_{B} $.
The normalized wavefunction coefficients read 
\begin{equation}
    (\alpha_n,\beta_n)_\pm^{\chi b}=
    \begin{cases}
    (\cos\frac{\theta_n}{2},\sin\frac{\theta_n}{2})  \\
    (-\sin\frac{\theta_n}{2},\cos\frac{\theta_n}{2})
    \end{cases}
\end{equation}
with $\theta_n=\arctan(p_n,q_n)$. Here, $\chi$ enters $q_n$ while $b$ does not affect the case of $n\geq1$, but their product will make a difference for LL$_0$ as we will have $\theta^{\chi b=+}_{0\pm}=\pi,0$ and $\theta^{\chi b=-}_{0\pm}=0,-\pi$, which assures the unique $0$th LL wavefunction and facilitates later calculations. Henceforth, we will also use LL$_n$ with $n\in\mathbb{Z}$ to denote all possible LLs, whose energies are instead expressed as
\begin{equation}\label{eq:LLs2}
    \varepsilon_{n}^{\chi b}=
    \begin{cases}
    \bar\varepsilon+\sgn(n)\,\kappa_{|n|} & n\neq0 \\
    \bar\varepsilon-\chi bm & n=0 
    \end{cases}.
\end{equation}

\subsection{Hall conductivity/conductance formula}
For each 2D $m(k_z)$-slice, the Kubo formula of general conductivity/conductance can be written as
\begin{equation}
    \sigma_{ij}=\frac{\ii e^2\hbar}{S}N_B\sum_{mn}\frac{f_{nm}}{\varepsilon_{mn}^2}v_{imn}v_{jnm},
\end{equation}
where $v_{imn}=\braket{\psi_m|v_i|\psi_n}$ is the matrix element between generic states LL$_m$ and LL$_n$, $f_{nm}=f_n-f_m$ with $f_n=f(\varepsilon_n)$ the Fermi distribution for LL$_n$, $\varepsilon_{mn}=\varepsilon_m-\varepsilon_n$, $N_B=Se|B|/h$ is the LL degeneracy and $S$ is the area in $ij$-plane. Then we find the Hall conductance
\begin{equation}\label{eq:sigmaH0}
     G_H=\frac{1}{2}(\sigma_{xy}-\sigma_{yx})=\frac{e^3|B|}{2\pi\hbar^2}\sum_{mn}A_{mn}
\end{equation}
where we suppress the $k_z$ dependence for brevity and $A_{mn}=B_{mn}C_{mn}$ with the property $A_{mn}=A_{nm}$, $B_{mn}=\frac{f_{mn}}{\varepsilon_{mn}^2}$ and
\begin{equation}\label{eq:C_def}
    C_{mn}=\hbar^2 \Im (v_{xmn}v_{ynm}) .
\end{equation}
We henceforth focus on the zero temperature case where $f_n=\Theta(\mu-\varepsilon_n)$ with $\mu$ the chemical potential.

\subsubsection{Velocity matrix elements}
Since the vector potential $\bA$ commutes with $\br$ and the velocity operator is generically defined as $\bv=\frac{\ii}{\hbar}[h(\bpi),\br]$, we can cast the velocity operator into $\bv=\frac{\partial h(\bk)}{\hbar\partial \bk}\big|_{\hbar\bk\rightarrow\bpi}$, leading to
\begin{equation}
\begin{split}
    \hbar v_i&=-t\pi_i\sigma_3+t_1( \delta_{i,x}\chi+ \delta_{i,y})\sigma_i=\begin{pmatrix}
       -\frac{d_i\bar v_{B} }{2a_i}(c^++e_ic^-) & t_1a_ie_ig_i \\
       t_1 a_ig_i & \frac{d_i\bar v_{B} }{2a_i}(c^++e_ic^-)
    \end{pmatrix}
\end{split}
\end{equation}
with $a_i=1,\ii$, $e_i=1,-1$, $f_i=1,\chi$, $g_i=\chi,1$, and $d_i=1,b$ for $i=x,y$ and $\bar v_{B} =4t_1 u_{B}/v_{B} $. Note that $b$ and $\chi$ respectively enter $v_y$ and $v_x$ only. 
We first write down the application of the velocity operator to a generic LL state
% \begin{equation}
%     \hbar v_i\psi_{ns}^b=
%     \begin{pmatrix}
%     (-\frac{\bar v_{B} }{2a_i}\alpha_{ns}\sqrt{n}+t_1a_ie_i\beta_{ns})\,
%     % \begin{cases}
%     %     \phi_n\\
%     %     \phi_{n-1}
%     % \end{cases}
%     \phi_{n+\bar b} - \frac{\bar v_{B} }{2a_i}\alpha_{ns}e_i\sqrt{n-b}\,
%     % \begin{cases}
%     %     \phi_{n-2}\\
%     %     \phi_{n+1}
%     % \end{cases}
%     \phi_{n-2-3\bar b} \\
%     \frac{\bar v_{B} }{2a_i}\beta_{ns}\sqrt{n+b}\,
%     % \begin{cases}
%     %     \phi_{n+1}\\
%     %     \phi_{n-2}
%     % \end{cases}
%     \phi_{n+1+3\bar b} + (\frac{\bar v_{B} }{2a_i}\beta_{ns}e_i\sqrt{n}+t_1a_i\alpha_{ns})\,
%     % \begin{cases}
%     %     \phi_{n-1}\\
%     %     \phi_{n}
%     % \end{cases}
%     \phi_{n-1-\bar b}
%     \end{pmatrix}
% \end{equation}
% with $\bar b=(b-1)/2$.
\begin{equation}
    \hbar v_i\psi_{ns}^{\chi b}=
    \begin{pmatrix}
    (-\frac{\bar v_{B} }{2a_i}\alpha_{ns}f_i\sqrt{n}+t_1a_ie_ig_i\beta_{ns})\,
    \phi_{n+\bar b} - \frac{\bar v_{B} }{2a_i}\alpha_{ns}e_if_i\sqrt{n-\chi b}\,
    \phi_{n-2-3\bar b} \\
    \frac{\bar v_{B} }{2a_i}\beta_{ns}f_i\sqrt{n+\chi b}\,
    \phi_{n+1+3\bar b} + (\frac{\bar v_{B} }{2a_i}\beta_{ns}e_if_i\sqrt{n}+t_1a_ig_i\alpha_{ns})\,
    \phi_{n-1-\bar b}
    \end{pmatrix}
\end{equation}
with $\bar b=(\chi b-1)/2$.
Then we find the nonzero matrix elements are 
% \begin{equation}\label{eq:v_matele}
% \begin{split}
%     % \braket{\psi_{n\pm1,s'}|v_i|\psi_{n,s}}^{b=+}&=
%     % \begin{cases}
%     %     \alpha_{n+1,s'}(-\frac{\bar v_{B} }{2a_i}\alpha_{ns}\sqrt{n}+t_1a_ie_i\beta_{ns}) + \frac{\bar v_{B} }{2a_i}\beta_{ns}\beta_{n+1,s'}\sqrt{n+1} \\
%     %     -\frac{\bar v_{B} }{2a_i}\alpha_{ns}\alpha_{n-1,s'}e_i\sqrt{n-1} + \beta_{n-1,s'}(\frac{\bar v_{B} }{2a_i}\beta_{ns}e_i\sqrt{n}+t_1a_i\alpha_{ns})
%     % \end{cases}\\
%     % \braket{\psi_{n\pm1,s'}|v_i|\psi_{n,s}}^{b=-}&=
%     % \begin{cases}
%     %     -\frac{\bar v_{B} }{2a_i}\alpha_{ns}\alpha_{n+1,s'}e_i\sqrt{n+1} + \beta_{n+1,s'}(\frac{\bar v_{B} }{2a_i}\beta_{ns}e_i\sqrt{n}+t_1a_i\alpha_{ns})\\
%     %     \alpha_{n-1,s'}(-\frac{\bar v_{B} }{2a_i}\alpha_{ns}\sqrt{n}+t_1a_ie_i\beta_{ns}) + \frac{\bar v_{B} }{2a_i}\beta_{ns}\beta_{n-1,s'}\sqrt{n-1} 
%     % \end{cases}\\
%     \braket{\psi_{n+ r,s'}|\hbar v_i|\psi_{n,s}}^{b}&=
%     \begin{cases}
%         \alpha_{n+b,s'}(-\frac{\bar v_{B} }{2a_i}\alpha_{ns}\sqrt{n}+t_1a_ie_i\beta_{ns}) + \frac{\bar v_{B} }{2a_i}\beta_{ns}\beta_{n+b,s'}\sqrt{n+b} & br=+\\
%         -\frac{\bar v_{B} }{2a_i}\alpha_{ns}\alpha_{n-b,s'}e_i\sqrt{n-b} + \beta_{n-b,s'}(\frac{\bar v_{B} }{2a_i}\beta_{ns}e_i\sqrt{n}+t_1a_i\alpha_{ns}) & br=-
%     \end{cases}
% \end{split}
% \end{equation}
\begin{equation}\label{eq:v_matele}
\begin{split}
    \braket{\psi_{n+ r,s'}|\hbar v_i|\psi_{n,s}}^{\chi b}&=
    \begin{cases}
        \alpha_{n+\chi b,s'}(-\frac{\bar v_{B} }{2a_i}\alpha_{ns}f_i\sqrt{n}+t_1a_ie_ig_i\beta_{ns}) + \frac{\bar v_{B} }{2a_i}\beta_{ns}\beta_{n+\chi b,s'}f_i\sqrt{n+\chi b} & \chi br=+\\
        -\frac{\bar v_{B} }{2a_i}\alpha_{ns}\alpha_{n-\chi b,s'}e_i\sqrt{n-\chi b} + \beta_{n-\chi b,s'}(\frac{\bar v_{B} }{2a_i}\beta_{ns}e_if_i\sqrt{n}+t_1a_ig_i\alpha_{ns}) & \chi br=-
    \end{cases}
\end{split}
\end{equation}
with $r=\pm1$.
In terms of the LL$_{n\in\mathbb{Z}}$ notation, Eq.~\eqref{eq:v_matele} suggests we consider the following two types of inter-LL matrix elements $v_{mn}$ with
\begin{equation}\label{eq:m_types}
    m=\begin{cases}
        n\pm1 &  G_H^+ \\
        -(n\pm1) &  G_H^-
    \end{cases},
\end{equation}
contributing to two different parts of $ G_H= G_H^++ G_H^-$ as indicated.

Now we calculate Eq.~\eqref{eq:C_def} for the two types in Eq.~\eqref{eq:m_types}, which is based on Eq.~\eqref{eq:v_matele}. 
For brevity, we mainly present the $\chi=1$ case and recover its dependence later when necessary.
We use the notation $\bar n=n+\frac{1}{2},\,\tilde{n}=\sgn(n)(n+\frac{1-\sgn(n)}{2})$.
For the first type in Eq.~\eqref{eq:m_types}, we find
\begin{equation}\label{eq:Ctype1}
\begin{split}
    C_{n+1,n}^{b}=-C_{n,n+1}^{b}&=\frac{\sgn(b\bar n)}{4}J^b_{\tilde{n}_b} 
\end{split}
\end{equation}
with 
\begin{equation}
\begin{split}
    J_n^+&=\left[ \bar v_{B} \sqrt{n}\cos{\frac{\theta_{n+1}'}{2}}\cos{\frac{\theta_{n}'}{2}} - \sin{\frac{\theta_{n}'}{2}} (\bar v_{B} \sqrt{n+1}\sin{\frac{\theta_{n+1}'}{2}}+2t_1\cos{\frac{\theta_{n+1}'}{2}}) \right]^2 \\
    J^-_n&=\left[ \bar v_{B} \sqrt{n+1}\cos{\frac{\theta_{n+1}}{2}}\cos{\frac{\theta_{n}}{2}} - \sin{\frac{\theta_{n+1}}{2}} (\bar v_{B} \sqrt{n}\sin{\frac{\theta_{n}}{2}}+2t_1\cos{\frac{\theta_{n}}{2}}) \right]^2 
\end{split}
\end{equation}
where $\theta'=\theta-\frac{\pi}{2}[\Theta(-n) \delta_{b,1}+ \delta_{n,-1} \delta_{b,-1}]$, $\theta^b_{0\sgn(\bar n)}$ is used and $\tilde{n}_+=\tilde{n}$ while $\tilde{n}_-$ means $n\rightarrow\sgn(n)n,n+1\rightarrow\sgn(n)(n+1)$.
For the second type in Eq.~\eqref{eq:m_types}, we find
\begin{equation}\label{eq:Ctype2}
\begin{split}
    C_{-(n+1),n}^{b=+}=C_{-n,(n+1)}^{b=-}&=\frac{\sgn(\bar n)}{4}K^+_{\tilde{n}} \\
    C_{-n,(n+1)}^{b=+}=C_{-(n+1),n}^{b=-}&=\frac{\sgn(-\bar n)}{4}K^-_{\tilde{n}} \\
    C_{-(n+1),n}^{b=+}+C_{-n,(n+1)}^{b=+}&=\frac{\sgn(-\bar n)}{8}K_{\tilde{n}}
\end{split}
\end{equation}
with 
\begin{equation}
\begin{split}
    K^+_n&=\left[ \bar v_{B} \sqrt{n+1}\cos{\frac{\theta_{n+1}}{2}}\sin{\frac{\theta_{n}}{2}} + \sin{\frac{\theta_{n+1}}{2}} (\bar v_{B} \sqrt{n}\cos{\frac{\theta_{n}}{2}}-2t_1\sin{\frac{\theta_{n}}{2}}) \right]^2 \\
    K^-_n&=\left[ \bar v_{B} \sqrt{n}\cos{\frac{\theta_{n+1}}{2}}\sin{\frac{\theta_{n}}{2}} + \cos{\frac{\theta_{n}}{2}} (\bar v_{B} \sqrt{n+1}\sin{\frac{\theta_{n+1}}{2}}+2t_1\cos{\frac{\theta_{n+1}}{2}}) \right]^2 \\
    K_n&= (\bar v_{B} ^2+4t_1^2)\cos{\theta_{n}}-(\bar v_{B} ^2-4t_1^2)\cos{\theta_{n+1}} + \bar v_{B} t_1(\sqrt{n}\sin{\theta_{n}}+\sqrt{n+1}\sin{\theta_{n+1}}) 
\end{split}
\end{equation}
where $\theta^+_{0\pm}$ is used in $K^\pm$.

\subsection{Hall conductance of reference state}
Note that there exists double counting in Eq.~\eqref{eq:m_types} by solving $n+1=-(n\pm1)$ and $n-1=-(n\pm1)$, leading respectively to the terms $A_{n+1,n}$ and $A_{n,n+1}$ for $n=0,-1$ in Eq.~\eqref{eq:sigmaH0}. We hence count them solely in $ G_H^+$. Assuming \textit{all} LL$_{n\leq n_0}$ are filled and otherwise empty, we have $f_{n\pm r}-f_n=-r \delta_{n,n_0+\frac{1-r}{2}}$ with $r=\pm1$ and can readily find 
\begin{equation}\label{eq:sigmaH+0}
     G_H^+=\frac{e^3|B|}{2\pi\hbar^2}\sum_n(A_{n+1,n}+A_{n-1,n})=\frac{e^3|B|}{2\pi\hbar^2}(A_{n_0+1,n_0}+A_{n_0,n_0+1}),%=\frac{e^3|B|}{2\pi\hbar^2}2A_{n_0+1,n_0},
\end{equation}
i.e., $ G_H^+$ is contributed by only one special pair of states LL$_{n_0}$ and LL$_{n_0+1}$. This is because $ G_H^+$ is a \textit{Fermi surface} contribution to the conductance.
On the other hand, for $ G_H^-$ that is the \textit{Fermi sea} contribution, its two involved energy denominators generally satisfy $\varepsilon_{-(n+1),n}=\varepsilon_{-n,n+1}$ unless $n=0,-1$.
Restricted to the above filling condition, the involved Fermi distributions differ as $f_{-(n+1),n}-f_{-n,n+1}=- \delta_{n,n_0}+ \delta_{n,-(n_0+1)}$, i.e., their difference can be neglected when $n_0=0,-1$ and hence $n=0,-1$ since such cases are accounted for in $ G_H^+$ as aforementioned. Therefore, when $n_0=0,-1$, we have the simplification
\begin{equation}\label{eq:sigmaH-0}
     G_H^-=\frac{e^3|B|}{2\pi\hbar^2}\sum_n(A_{-(n+1),n}+A_{-n,n+1})=\frac{e^3|B|}{2\pi\hbar^2}\sum_{n\neq0,-1} B_{-(n+1),n}(C_{-(n+1),n}+C_{-n,n+1}).
\end{equation}
Lastly, it is crucial to note that the above simplest filling situation specified by some $n_0$ is \textit{rarely} the case except when the chemical potential lies in the `gap' between LL$_{n>0}$ and LL$_{n<0}$ (chiral LL$_0$ always makes the system overall gapless). This is because, with the variable mass $m(k_z)$ in Eq.~\eqref{eq:h} and Eq.~\eqref{eq:LLs}, the \textit{ordering} of LLs will reshuffle in complex ways when $tm>0$, i.e., when we have topologically nontrivial AHE as indicated later by Eq.~\eqref{eq:barsigmaHchi}. 
Note also that such reshuffling happens respectively \textit{within} and does \textit{not mix} electron-like LL$_{n>0}$ and hole-like LL$_{n\leq0}$ when $\sgn(t\chi b)=+$ (electron-like LL$_{n\geq0}$ and hole-like LL$_{n<0}$ when $\sgn(t\chi b)=-$). This is because the `gap' between LL$_{n>0}$ and LL$_{n<0}$ never closes except the gapless LL$_0$ switches its chirality by reversing $\chi b$.

Given these properties of $ G_H^\pm$ and LLs, it is convenient to take advantage of the only two possible special fillings $n_0=0,-1$ as the reference state and then consider the change $\Delta G_H$ for any arbitrary filling. Note that the meaning of $n_0$ is that \textit{all} LL$_{n\leq n_0}$ are filled and otherwise empty. Importantly, according to Eq.~\eqref{eq:LLs}, this reference state is not a unique state; instead, it encompasses four possible combinations concerning the configuration of chiral LL$_0$ and the sign of mass. Summarized in Table~\ref{table1}, we introduce a $\chi b,\sgn(m)$-dependent $n_0$ as $n_0^{\chi b\gtrless}=\frac{-1+\sgn(m)\chi b}{2}$. It does \textit{not} depend on $\sgn(t)$ since the chirality of LL$_0$ is only controlled by $\chi b$, which is also consistent with Eq.~\eqref{eq:LLsym1}.
\begin{table}%The best place to locate the table environment is directly after its first reference in text
%\begin{ruledtabular}
\begin{tabular}{c|c|c}
%\cline{2-5}
$\chi b$ & $+$ & $-$\\
%\cline{2-4}
\hline
$m>0,n_0^{\chi b>}$ & $0$ & $-1$\\
\hline
$m<0,n_0^{\chi b<}$ & $-1$ & $0$\\
%\hline
\end{tabular}
%\end{ruledtabular}
\caption{Highest filled LL $n_0^{\chi b\gtrless}$ for \textit{arbitrary} Hamiltonian parameters including $t$; chemical potential only crosses chiral LL$_0$ in the $(m,\varepsilon)$-plane. It apparently depends on the signs of mass, magnetic field, and chirality but not that of $t$.}\label{table1}
\end{table}

Let's now evaluate the Hall conductance $\bar{G}_H$ for the reference state with $n_0=0,-1$. We use the notation in Eq.~\eqref{eq:LLs} and further denote $g_n^\pm=p_n(v_{B} ^2\pm4 u_{B}^2)+4n u_{B} v_{B} ^2$ and $Z_n=\frac{p_n}{\kappa_n}-\frac{p_{n+1}}{\kappa_{n+1}}$, from which we establish the identity below for $n\in\mathbb{N}$
\begin{equation}\label{eq:Z_n}
\begin{split}
    \frac{\frac{g_n^+}{\kappa_n}+\frac{g_{n+1}^-}{\kappa_{n+1}}}{(\kappa_n+\kappa_{n+1})^2}&=\frac{(\frac{g_n^+}{\kappa_n}+\frac{g_{n+1}^-}{\kappa_{n+1}})(\kappa_{n+1}-\kappa_{n})}{(\kappa_{n+1}+\kappa_{n})(\kappa_{n+1}^2-\kappa_{n}^2)} = \frac{(g_n^+\kappa_{n+1}^2-g_{n+1}^-\kappa_n^2)+\kappa_{n}\kappa_{n+1}(g_{n+1}^--g_{n}^+)}{\kappa_{n}\kappa_{n+1}(\kappa_{n}+\kappa_{n+1})(\kappa_{n+1}^2-\kappa_{n}^2)} \\
    &= \frac{(v_{B} ^2m-2 u_{B} p_np_{n+1})+2 u_{B}\kappa_{n}\kappa_{n+1}}{\kappa_{n}\kappa_{n+1}(\kappa_{n}+\kappa_{n+1})} = \frac{[(v_{B} ^2m-2 u_{B} p_np_{n+1})+2 u_{B}\kappa_{n}\kappa_{n+1}](\kappa_{n+1}-\kappa_{n})}{\kappa_{n}\kappa_{n+1}(\kappa_{n+1}^2-\kappa_{n}^2)} \\
    &= \frac{m(\kappa_{n+1}-\kappa_{n})+2 u_{B}[(n+1)\kappa_n-n\kappa_{n+1}]}{\kappa_{n}\kappa_{n+1}} = Z_n.
\end{split}
\end{equation}
From Eq.~\eqref{eq:sigmaH+0} and Eq.~\eqref{eq:Ctype1}, we find
\begin{equation}\label{eq:sigmaH+1}
\begin{split}
    \bar{G}_H^+&=\frac{e^3|B|}{2\pi\hbar^2}2A_{n_0+1,n_0} =\frac{e^3|B|}{2\pi\hbar^2}2
    \begin{cases}
        \frac{-1}{(m+\kappa_1)^2}\frac{1}{4}(\bar v_{B} \sin{\frac{\theta_1}{2}}+2t_1\cos{\frac{\theta_1}{2}})^2 & m>0,n_0^{b>}=0,-1 \\
        \frac{-1}{(-m+\kappa_1)^2}\frac{-1}{4}(\bar v_{B} \cos{\frac{\theta_1}{2}}-2t_1\sin{\frac{\theta_1}{2}})^2 & m<0,n_0^{b<}=-1,0
    \end{cases} \\
    &=\frac{e^3|B|}{2\pi\hbar^2}\frac{-1}{4}\frac{4t_1^2}{v_{B} ^2}
    \begin{cases}
        \frac{1}{(m+\kappa_1)^2}(\frac{g_0^+}{m}+\frac{g_1^-}{\kappa_1}) & m>0,n_0^{b>}=0,-1 \\
        \frac{1}{(-m+\kappa_1)^2}(\frac{g_0^+}{-m}+\frac{g_1^-}{\kappa_1}) & m<0,n_0^{b<}=-1,0
    \end{cases} \\
    &=\frac{e^2}{h}\frac{-1}{2}\frac{\frac{g_0^+}{\kappa_0}+\frac{g_1^-}{\kappa_1}}{(\kappa_0+\kappa_1)^2} =\frac{e^2}{h}\frac{-1}{2}Z_0
\end{split}
\end{equation}
where the two cases follow Table~\ref{table1}.
From Eq.~\eqref{eq:sigmaH-0} and Eq.~\eqref{eq:Ctype2}, we find
\begin{equation}\label{eq:sigmaH-1}
\begin{split}
    \bar{G}_H^-&=\frac{e^3|B|}{2\pi\hbar^2}\sum_{n\neq0,-1}\frac{\sgn(\bar n)}{\varepsilon_{-(n+1),n}^2}\frac{\sgn(-\bar n)}{8}K_{\tilde{n}} \\
    &=\frac{e^3|B|}{2\pi\hbar^2}\frac{-1}{8}\left[\sum_{n=1}^{+\infty}\frac{K_n}{\varepsilon_{-(n+1),n}^2}+\sum_{n=-2}^{-\infty}\frac{K_{-(n+1)}}{\varepsilon_{-(n+1),n}^2}\right] =\frac{e^3|B|}{2\pi\hbar^2}\frac{-1}{8}\left[\sum_{n=1}^{+\infty}\frac{K_n}{\varepsilon_{-(n+1),n}^2}+\sum_{n=1}^{+\infty}\frac{K_{n}}{\varepsilon_{n,-(n+1)}^2}\right]  \\
    &=\frac{e^3|B|}{2\pi\hbar^2}\frac{-1}{4}\sum_{n=1}^{+\infty}\frac{K_n}{\varepsilon_{-(n+1),n}^2}  =\frac{e^2}{h}\frac{-1}{2}\sum_{n=1}^{+\infty}\frac{\frac{g_n^+}{\kappa_n}+\frac{g_{n+1}^-}{\kappa_{n+1}}}{(\kappa_n+\kappa_{n+1})^2} =\frac{e^2}{h}\frac{-1}{2}\sum_{n=1}^{+\infty}Z_n.
\end{split}
\end{equation}
In the last step of Eq.~\eqref{eq:sigmaH+1} and Eq.~\eqref{eq:sigmaH-1}, we make use of Eq.~\eqref{eq:Z_n}. 
Then we can calculate the full Hall conductance for the reference state
\begin{equation}\label{eq:barsigmaH}
\begin{split}
    \bar{G}_H&=\bar{G}_H^++\bar{G}_H^- = \frac{e^2}{h}\frac{-1}{2}\sum_{n=0}^{+\infty}Z_n = \frac{e^2}{h}\frac{-1}{2}(\frac{p_0}{\kappa_0}-\frac{p_{n+1}}{\kappa_{n+1}}\big|_{n\rightarrow\infty}) = \frac{e^2}{h}\frac{-1}{2}[\sgn(m)+\sgn( u_{B})].
\end{split}
\end{equation}
For the reference state, i.e., as long as the chemical potential is within the `gap' excluding LL$_0$, $\bar{G}_H$ is quantized to be either 0 or -1 in units of $\frac{e^2}{h}$ \textit{regardless} of the magnetic field $B$. Given $ u_{B}>0$ throughout this study, the topologically (non)trivial case bears negative (positive) mass $m$. When $|B|=0$ we have Eq.~\eqref{eq:barsigmaH} reduced to
\begin{equation}\label{eq:barsigmaH_B=0}
    \bar{G}_H= \frac{e^2}{h}\frac{-1}{2}[\sgn(m)+\sgn(t)],
\end{equation}
as can be seen from the $|B|\rightarrow0$ limit or otherwise directly calculated. Also, note that Eq.~\eqref{eq:barsigmaH} becomes 
\begin{equation}\label{eq:barsigmaHchi}
\begin{split}
    \bar{G}_H&= \frac{e^2}{h}\frac{-\chi}{2}[\sgn(m)+\sgn(t)]
\end{split}
\end{equation}
once we recover the chirality dependence. This is physically expected as this AHE contribution is purely due to the intrinsic chirality $\chi$ of the WSM system.

\subsection{Conductance change at arbitrary filling}
\subsubsection{Single-LL filling}
Now we turn to find the change $\Delta G_H$ for arbitrary filling with respect to the reference state of Table~\ref{table1} with conductance $\bar{G}_H$ in
Eq.~\eqref{eq:barsigmaH}. It is crucial to note that the filling procedure is associative and commutative. Thus, we can study the effect of `single-LL' fillings below as the building blocks for any arbitrary fillings, since $\Delta G_H$ is an \textit{additive} map of such single-LL fillings. As aforementioned, an arbitrary filling is generally \textit{not} simply labelled by an $n_0$ for the top filled LL, because of the LL reshuffling. Below, we will consider three representative cases related to LL$_{n\geq0}$ in order to construct the general result, the situation with the filling of LL$_{n<0}$ modified can be similarly and straightforwardly expected.

\begin{enumerate}
\item Fill LL$_{n>1}$

We change from $f_n=0$ to $f_n=1$ with respect to the reference state in Table~\ref{table1}. We find the corresponding change in conductance
\begin{equation}
\begin{split}
    \Delta G_H^b&=\frac{e^3|B|}{2\pi\hbar^2}\left[ (A_{n+1,n}+A_{n,n+1})_{a} - (A_{n+1,n}+A_{n,n+1})_{n\rightarrow n-1,a'} \right. \\
    &\left. - (A_{-(n+1),n}+A_{n,-(n+1)})_{r} - (A_{-n,n+1}+A_{n+1,-n})_{n\rightarrow n-1,r} \right]^b \\
    &= \frac{e^3|B|}{2\pi\hbar^2}2\left[ (A_{n+1,n;a} - A_{-(n+1),n;r} ) + (-A_{n,n-1;a'}  - A_{-(n-1),n;r}) \right]^b ,
\end{split}
\end{equation}
where subscript $a/r$ respectively means adding new/removing old contributions due to the modified filling. While both $a,a'$ account for adding new terms from filling a particular LL$_m$, type-$a$ comes from coupling with LL$_{n>m}$ and type-$a'$ comes from coupling with LL$_{n<m}$; hence the negative sign of $a'$ term due to the reversed difference of Fermi functions, i.e., $f_{n+1}-f_n=-(f_{n}-f_{n-1})$. Importantly, this sign difference assists in cancelling various terms when filling consecutive LLs, thus enabling the \textit{additivity} of $\Delta G_H$ aforementioned. 
Denoting $(A_{n+1,n;a} - A_{-(n+1),n;r} )^b=\frac{t_1^2}{2v_{B} ^2}X_n^b,\,(-A_{n,n-1;a'}  - A_{-(n-1),n;r})^b=\frac{t_1^2}{2v_{B} ^2}Y_n^b$, we find from Eq.~\eqref{eq:Ctype1} and Eq.~\eqref{eq:Ctype2} that
\begin{equation}\label{eq:X_nY_n}
\begin{split}
    X_n^b&=\frac{(\kappa_n^2-\kappa_{n+1}^2)D_{n+}^b+\kappa_n\kappa_{n+1}(E_{n+}^b+F_{n+}^b)}{(\kappa_n^2-\kappa_{n+1}^2)^2} = \frac{p_n-b(2n+1)\kappa_n}{\kappa_n}\\
    Y_n^b&=\frac{(\kappa_n^2-\kappa_{n-1}^2)D_{n-}^b+\kappa_n\kappa_{n-1}(E_{n-}^b+F_{n-}^b)}{(\kappa_n^2-\kappa_{n-1}^2)^2} = -\frac{p_n-b(2n-1)\kappa_n}{\kappa_n}
\end{split}
\end{equation}
where 
\begin{equation}
\begin{split}
    D_{n\pm}^b&=\cos{\theta_n}(v_{B} ^2\pm4 u_{B}^2)4\sqrt{n}v_{B}  u_{B}\sin{\theta_n} \mp b[4(2n\pm1) u_{B}^2+v_{B} ^2] \\
    % E_{n\pm}^{b=+}&=\cos{\theta_{n\pm1}}\{\cos{\theta_{n}}[v_{B} ^2-4(2n\pm1) u_{B}^2] + 4\sqrt{n}v_{B}  u_{B}\sin{\theta_n} \mp v_{B} ^2 + 4 u_{B}^2 \} \\
    % E_{n\pm}^{b=-}&=\mp\cos{\theta_{n\pm1}}\{\cos{\theta_{n}}[v_{B} ^2-4(2n\pm1) u_{B}^2] + 4\sqrt{n}v_{B}  u_{B}\sin{\theta_n} \pm v_{B} ^2 - 4 u_{B}^2 \} \\
    E_{n\pm}^{b}&=\bar b_\pm\cos{\theta_{n\pm1}}\{\cos{\theta_{n}}[v_{B} ^2-4(2n\pm1) u_{B}^2] + 4\sqrt{n}v_{B}  u_{B}\sin{\theta_n} +b(\mp v_{B} ^2 + 4 u_{B}^2) \} \\
    F_{n\pm}^{b}&=\bar b_\pm4\sqrt{n\pm1} u_{B}\sin{\theta_{n\pm1}}[v_{B} (\cos{\theta_n}\mp b) + 2\sqrt{n} u_{B}\sin{\theta_n}]
\end{split}
\end{equation}
with $\bar b_\pm=1 \delta_{b,1}+(\mp1) \delta_{b,-1}$.
Therefore, we obtain 
\begin{equation}
    \Delta G_H^b= \frac{e^3|B|}{2\pi\hbar^2}\frac{2t_1^2}{2v_{B} ^2}(X_n+Y_n)^b=-b\frac{e^2}{h}.
\end{equation}

\item Fill LL$_1$

We change from $f_1=0$ to $f_1=1$ with respect to the reference state in Table~\ref{table1}. We find the change in conductance
\begin{equation}
    \Delta G_H^b=\frac{e^3|B|}{2\pi\hbar^2}\left[ (A_{1,2}+A_{2,1})_{a} - (A_{1,-2}+A_{-2,1})_{r} + 
        \begin{cases}
            -(A_{1,0}+A_{0,1})_r & n_0=0 \\
            -(A_{1,0}+A_{0,1})_{a'} & n_0=-1
        \end{cases}
        \right]^b.
\end{equation}
Here, since $(A_{1,0}+A_{0,1})_r =(A_{1,0}+A_{0,1})_{a'}$ we can compactly use $n_0=0$ only to represent it, which actually means $n_0^{b=+,>}=0$ or $n_0^{b=-,<}=0$. Then we find 
\begin{equation}
\begin{split}
    \Delta G_H^b&=\frac{e^3|B|}{2\pi\hbar^2}\left[ (A_{1,2}+A_{2,1})_{a} - (A_{1,-2}+A_{-2,1})_{r} - (A_{1,0}+A_{0,1})_r \right]^b \\
    &= \frac{e^3|B|}{2\pi\hbar^2}2\left[ \left(A_{2,1;a} - A_{-2,1;r}\right) - A_{1,0;r} \right]^b \\
    &= \frac{e^3|B|}{2\pi\hbar^2}2\frac{t_1^2}{2v_{B} ^2}[X_1^b-(\frac{p_1}{\kappa_1}-\frac{p_0}{bm})] = -b \frac{e^2}{h}
\end{split}
\end{equation}
where we use Eq.~\eqref{eq:sigmaH+1} and Eq.~\eqref{eq:X_nY_n}.

\item Move $\mu$ within the `gap', i.e., fill or empty LL$_0$

\begin{itemize}
    \item $m>0$
    
    We start from $m>0$ with $n_0^{b>}=0,-1$ and move down (up) to empty (fill) LL$_0$ when $b=+$ ($b=-$), i.e., realizing $\begin{cases}
        f_{n\geq0}=0,f_{n<0}=1 & b=+ \\
        f_{n>0}=0,f_{n\leq0}=1 & b=- 
    \end{cases}$.
    Taking $b=+$ as an example, we note that although $\varepsilon_{n\leq0}$ may cross and reorder with increasing $m$, we can think of small $m$ region where crossing is yet to occur and hence the relevance of this discussion.
    Considering the change in filling, we find
    \begin{equation}
        \Delta G_H^b=\frac{e^3|B|}{2\pi\hbar^2}
        \begin{cases}
            \left[ (A_{-1,0}+A_{0,-1})_a - (A_{1,0}+A_{0,1})_r \right]^{b=+} \\
            \left[ (A_{1,0}+A_{0,1})_a - (A_{-1,0}+A_{0,-1})_r \right]^{b=-}
        \end{cases}
    \end{equation}
    and from Eq.~\eqref{eq:Ctype1} and Eq.~\eqref{eq:Ctype2} we have 
    \begin{equation}
    \begin{split}
        (A_{\pm1,0}+A_{0,\pm1})_r^\pm=2A_{0,1}&=2\frac{1}{(m+\kappa_1)^2}\frac{-1}{4}(\bar v_{B} \sin{\frac{\theta_1}{2}}+2t_1\cos{\frac{\theta_1}{2}})^2 \\
        (A_{\pm1,0}+A_{0,\pm1})_a^\mp=2A_{-1,0}&=2\frac{1}{(-m+\kappa_1)^2}\frac{1}{4}(\bar v_{B} \cos{\frac{\theta_1}{2}}-2t_1\sin{\frac{\theta_1}{2}})^2.
    \end{split}
    \end{equation}
    Reusing the result in Eq.~\eqref{eq:sigmaH+1}, we obtain 
    \begin{equation}
        \Delta G_H^b=\frac{e^2}{h}\frac{-1}{2}[(\frac{p_0}{-m}-\frac{p_1}{\kappa_1})-(\frac{p_0}{m}-\frac{p_1}{\kappa_1})]=\frac{e^2}{h}.
    \end{equation}
    
    \item $m<0$

    We start from $m<0$ with $n_0^{b<}=-1,0$ and move up (down) to fill (empty) LL$_0$ when $b=+$ ($b=-$), i.e., realizing $\begin{cases}
        f_{n>0}=0,f_{n\leq0}=1 & b=+ \\
        f_{n\geq0}=0,f_{n<0}=1 & b=- 
    \end{cases}$. Similarly, we find
    \begin{equation}
        \Delta G_H^b=\frac{e^3|B|}{2\pi\hbar^2}
        \begin{cases}
            \left[ (A_{-1,0}+A_{0,-1})_a - (A_{1,0}+A_{0,1})_r \right]^{b=-} \\
            \left[ (A_{1,0}+A_{0,1})_a - (A_{-1,0}+A_{0,-1})_r \right]^{b=+}
        \end{cases}=-\frac{e^2}{h},
    \end{equation}
    which is opposite to the $m>0$ case.

\end{itemize}
Note that the independence of $\Delta G_H^b$ on $b$ in both formulae is merely apparent since we fix to specific filling actions respectively.

\end{enumerate}
In summary, we find that, with respect to the reference state, filling any one extra LL entails the quantized change $\Delta G_H^b=-b\frac{e^2}{h}$ and the opposite for emptying it, i.e.,
\begin{equation}\label{eq:Deltasigma_rule}
\Delta G_H^b=
    \begin{cases}
        -b\frac{e^2}{h} & \textrm{fill one extra LL} \\
        b\frac{e^2}{h} & \textrm{empty one extra LL}
    \end{cases}
\end{equation}
for \textit{any} $t,\chi$.
This is consistent with the physical picture of the QHE where each LL carries its edge channel responsible for the quantized conductance. It is worth noting that, in contrast to Eq.~\eqref{eq:barsigmaHchi}, even with the $\chi$-dependence recovered this conductance change remains \textit{independent} of $\chi$, because such a contribution, in the same manner as the normal Hall effect and QHE, is the purely field-induced Hall effect and independent of the system chirality.

\subsubsection{General analytic formula}
Now we need to find a general formula of $\Delta G_H^b$ for any arbitrary chemical potential $\mu$. The purpose is to count how many LLs are filled (empty) above (below) the `gap' and each of them contributes $-b\frac{e^2}{h}$ ($b\frac{e^2}{h}$) as per Eq.~\eqref{eq:Deltasigma_rule}. Instead of brute-force counting with an intractable infinite sum, we consider from Eq.~\eqref{eq:LLs} the generic equation of variable $n$
\begin{equation}
    \mu=\varepsilon_n
\end{equation}
and relax the integer $n$ such that $n\in\mathbb{R}$. The solution is 
\begin{equation}
    n_\pm=\frac{4m u_{B}-v_{B} ^2\pm\sqrt{\Delta}}{8 u_{B}^2}
\end{equation}
with the determinant $\Delta=v_{B} ^2(v_{B} ^2-8m u_{B})+16 u_{B}^2(\mu-\bar\varepsilon)^2$ with $\bar\varepsilon=\varepsilon_0(m=0)=\chi b u_{B}$. 
Then we define the following two key quantities
\begin{equation}\label{eq:n_LR}
    n_L=\max{[-1,\lfloor n_+\rfloor]},\; n_R=\max{[\xi ,\lceil n_-\rceil]}
\end{equation}
where $\lfloor x \rfloor$ ($\lceil x \rceil$) denotes the greatest (smallest) integer less (greater) than or equal to $x$ and $\xi =\frac{1+\sgn(m)\chi b\nu }{2}$ equals to $0$ or $1$ dependent on whether it is in the electron-like or hole-like region, i.e., $\nu =\sgn(\mu-\bar\varepsilon)=\pm1$ respectively. 
Geometrically in the $(m,\varepsilon)$-plane, an electron (hole) like LL lies below (above) $\mu$ if and only if the point $(m,\mu)$ under our inspection is located in between two crossing points, if exist at all, of the LL and $\varepsilon=\mu$.
Physically, $n_L$ ($n_R$) is the highest index of the LL with its left (right) crossing point to the left (right) of the point $(m,\mu)$. Special care is necessary for the gapless chiral LL$_0$ since it always crosses $\varepsilon=\mu$ once, which is taken into account in the minima $-1,\xi $ set in Eq.~\eqref{eq:n_LR}.
Finally, we can write down the following concise analytical formula
\begin{equation}\label{eq:DeltasigmaH}
    \Delta G_H=\frac{e^2}{h}(-b\nu )\Theta(\Delta)
        \Theta(n_L-n_R) \,(n_L-n_R+1) .
\end{equation}
% \red{\begin{equation}\label{eq:DeltasigmaH_old}
%     \textrm{(valid for $t>0$) } \Delta G_H=\frac{e^2}{h}(-b\nu )\Theta(\Delta)
%     \begin{cases}
%         \Theta(n_L-n_R) \,(n_L-n_R+1) & m>0 \\
%         \Theta(n_L) \,(n_L+\xi ) & m\leq0
%     \end{cases}.
% \end{equation}}
The factor $(-b\nu )$ directly comes from the quantized conductance from each LL we found above and accounts for the opposite contribution of filled or empty LLs via $\nu $. $\Theta(\Delta)$ means no contribution when there is completely no crossing between the chemical potential and LL$_{n\neq0}$, e.g. when it is within a certain region of the `gap' where even LL$_0$ does not contribute.
Then we exclude the redundant solutions through the rest $\Theta$ step function. %$m>0$ case depends on both $n_L,n_R$ because the LL reshuffling happens in this region; $m\leq0$ case respects the simple LL ordering and hence can be easily characterized by $n_L$ only.

\subsection{General formula of 3D Hall conductance}\label{App:3Dformula}

The 3D conductance of the whole system reads $G_H=\sum_{k_z} G_H(m(k_z))$ with the 2D Hall conductance $ G_H=\bar{G}_H + \Delta G_H$ combining Eq.~\eqref{eq:barsigmaH} and Eq.~\eqref{eq:DeltasigmaH}, where we now recover the $k_z$-dependence of the mass $m$. For the WSM system with $L_z$ the thickness along $z$-direction, we make use of the Dirichlet boundary condition per the infinite wall problem, leading to allowed discrete $k_z$ values 
\begin{equation}\label{eq:k_i}
    k_i=\frac{i\pi}{L_z}
\end{equation}
with $i\in\mathbb{N}^+$. An important difference from the $B=0$ WSM, where the AHE is only from the topologically nontrivial $m>0$ region between the WPs, lies in that the $m<0$ region outside the WPs ($m=0$) is \textit{also} able to contribute. Intuitively, the magnetic field turns such an originally topologically trivial region nontrivial via the LL formation. However, such contribution is only up to a certain negative \textit{lower bound} $m^*<0$, because for negatively large enough $m$ eventually all LLs return to the configuration of our reference state according to Eq.~\eqref{eq:LLs}. Such a state means that all electron (hole) like LLs are above (below) $\mu$, i.e., $\mu$ is between LL$_0$ and LL$_{-\chi b}$ since LLs are trivially ordered when $m<0$ as aforementioned. 
This is expected in realistic systems, since there always should exist a lower bound $m^*$ of $m(k_z)$ such that $ G_H(m<m^*)\equiv0$, which naturally stops the apparently infinite momentum summation. %and, interestingly, implies that Hall responses are contributed even outside the WPs ($m=0$), in contrast to the AHE. 
Based on Table~\ref{table1}, we accordingly find two constraints when $m<0$ so as to have finite $\Delta G_H$. 
\begin{itemize}
\item LL$_0$
\begin{equation}
\begin{cases}
    \mu>\varepsilon_0 & \textrm{LL$_0$ filled} \quad \chi b=+ \\
    \mu<\varepsilon_0 & \textrm{LL$_0$ empty} \quad \chi b=-
\end{cases}        
\implies m>m_1^*= u_{B}-\chi b\mu
\end{equation}

\item LL$_{-\chi b}$
\begin{equation}
\begin{cases}
    \mu<\varepsilon_{-\chi b} & \textrm{LL$_{-1}$ empty} \quad \chi b=+ \\
    \mu>\varepsilon_{-\chi b} & \textrm{LL$_1$ filled} \quad \chi b=-
\end{cases}        
\implies m>m_2^*=2 u_{B}-\sqrt{\mycomment{( u_{B}-\chi b\mu)^2}(\mu-\bar\varepsilon)^2-v_{B} ^2}
\end{equation}

\end{itemize}
To account also for the $t<0$ case, we note that the foregoing nonpositive lower mass bound becomes a nonnegative \textit{upper bound}. Making use of Eq.~\eqref{eq:LLsym1}, we find the compact form below for the same two constraints
\begin{equation}
\sgn(t\chi b)\mu>\sgn(t\chi b)\varepsilon_0 
\implies \sgn(t) m>m_1^*=\sgn(t)( u_{B}-\chi b\mu)
\end{equation}
and
\begin{equation}
\sgn(t\chi b)\mu<\sgn(t\chi b)\varepsilon_{-\sgn(t\chi b)} 
\implies \sgn(t) m>m_2^*=2 u_{B}\sgn(t)-\sqrt{\mycomment{( u_{B}-\chi b\mu)^2}(\mu-\bar\varepsilon)^2-v_{B} ^2}.
\end{equation}
Then we reach the generic mass bound 
\begin{equation}\label{eq:m*}
    \sgn(t)m^*=\min{[0,m_1^*,m_2^*]}
\end{equation}
where $0$ is included to account for the case when both $m_{1,2}^*>0$ and $m_2^*$ is neglected when it does not exist. 
We can now write down the final expression of the 3D conductance
\begin{equation}\label{eq:G_H}
    G_H=\sum_{i=1}^{i^*}[\bar{G}_H(m(k_i)) + \Delta G_H(m(k_i))],
\end{equation}
where 
\begin{equation}\label{eq:i*}
    i^*=\left\lfloor \frac{L_z}{\pi}\sqrt{\frac{2}{t_0}(m_0-m^*)} \right\rfloor
\end{equation}
follows from Eq.~\eqref{eq:m*}. Note that Eq.~\eqref{eq:i*} is invariant under the mappings of reversing $\mu,b$ or $m_0,t_0,t,b$ or $\chi,b$. This is used to prove Eq.~\eqref{eq:symm} and Eq.~\eqref{eq:symm2}.

\section{Symmetry relations}\label{App:symmetry}

We point out below a set of nontrivial symmetries relations held by the Hamiltonian, the LLs, and, especially, the conductance formula Eq.~\eqref{eq:G_H}.

We briefly comment on the symmetry of our Hamiltonians Eq.~\eqref{eq:H}, Eq.~\eqref{eq:h}, and Eq.~\eqref{eq:h^b}.  Firstly, in the original lattice Hamiltonian Eq.~\eqref{eq:H}, a symmetry $\mathcal{S}$, expressed as
    \begin{equation}
        \sigma_zH(\bk)\sigma_z^{-1}=H(-\bk),
    \end{equation}
relates the two WPs at $(0,0,\pm k_w)$. When $\sigma$ stands for real spin, $\mathcal{S}$ is a combined symmetry as $\mathcal{S}=\mathcal{I}\mathcal{R}_{z}^\pi$ with $\mathcal{I}$ the inversion and $\mathcal{R}_{z}^\pi$ the spin $\pi$-rotation along $z$. %as $\mathcal{S}=\mathcal{I}_{xy}\mathcal{M}_{z}$ with $\mathcal{I}_{xy}$ the in-plane inversion and $\mathcal{M}_{z}$ the mirror operation in $z$. 
When $\sigma$ stands for pseudospin, $\mathcal{S}$ could mean inversion $\mathcal{I}$ if the two orbitals are parity-even and parity-odd, e.g., $s$ and $p$ orbitals.
Without magnetic field, Eq.~\eqref{eq:h} preserves %the inversion $\mathcal{I}=\sigma_z$ and 
the particle-hole symmetry $\mathcal{P}=\sigma_x K$ with complex conjugation $K$, as is seen from %$\sigma_zh(-\bk)\sigma_z^{-1}=h(\bk)$ and 
$\sigma_xh^*(-\bk)\sigma_x^{-1}=-h(\bk)$. However, it no longer holds for Eq.~\eqref{eq:h^b} under the magnetic field and/or with the chemical potential term $-\mu\sigma_0$ added, which actually leads to the relation $\sigma_xh^*(B,\mu)\sigma_x^{-1}=-h(-B,-\mu)$ because $\mathcal{P}$ reverses momentum but preserves position and hence the vector potential, i.e., $\mathcal{P}\bpi(B)\mathcal{P}^{-1}=-\bpi(-B)$. Therefore, we have 
\begin{equation}\label{eq:PHsymm}
    \mathcal{P}\mathcal{H}(B,\mu)\mathcal{P}^{-1}=\mathcal{H}(-B,-\mu)
\end{equation}
for the full Hamiltonian $\mathcal{H}$. This immediately entails a modified $\mathcal{P}$-symmetry
\begin{equation}\label{eq:LL_PHsymm}
    \varepsilon_n(B,\mu)=-\varepsilon_{-n}(-B,-\mu)
\end{equation}
held by the LLs of Eq.~\eqref{eq:LLs} when $\mu$ is taken into account.

\subsection{Particle-hole relation}
The first symmetry held by Eq.~\eqref{eq:G_H} reads
\begin{equation}\label{eq:symm}
    G_H(B,\mu)=G_H(-B,-\mu).%=-G_H(-\chi ,-B,\mu).
\end{equation}
Note that $G_H(B,\mu)\neq G_H(-B,\mu)$ in general because the $\mathcal{T}$-breaking contribution exists in the first place even without finite $B$; such offset contribution is then compensated by reversing $\mu$ at once. %Also note that the LLs break particle-hole symmetry.
This symmetry can be proved as follows. i) For the part from the reference state, $\bar{G}_H$ does not depend on $B,\mu$. Then only the second part $\Delta G_H=\sum_{i=1}^{i^*} \Delta G_H(m(k_i))$ matters. 
ii) Given any $m$, $\Delta G_H(B,\mu,m)$ in Eq.~\eqref{eq:DeltasigmaH} is a function of $b\mu$ and $b\nu =\sgn(b\mu-\chi  u_{B})$; hence we also have $\Delta G_H(B,\mu,m)=\Delta G_H(-B,-\mu,m)$. 
iii) $k_i$ is independent of $B,\mu$ while $i^*$ depends. However, $i^*(B,\mu)=i^*(|B|,b\mu)$; therefore, we have $i^*(B,\mu)=i^*(-B,-\mu)$.

\begin{table}%The best place to locate the table environment is directly after its first reference in text
%\begin{ruledtabular}
\begin{tabular}{c|c|c|c}
%\cline{2-5}
\multicolumn{2}{c|}{system I} & \multicolumn{2}{c}{system II}\\
%\cline{2-4}
\hline
$\varepsilon_n(B,\mu)>0$ & LL$_n$ empty & $\varepsilon_{-n}(-B,-\mu)<0$ & LL$_{-n}$ filled \\
\hline
$\varepsilon_n(B,\mu)<0$ & LL$_n$ filled & $\varepsilon_{-n}(-B,-\mu)>0$ & LL$_{-n}$ empty\\
%\hline
\end{tabular}
%\end{ruledtabular}
\caption{Correspondence of LL occupancy between systems with $B,\mu$ and $-B,-\mu$.}\label{table2}
\end{table}

Physically, Eq.~\eqref{eq:symm} is a consequence of combining the WSM topology and the particle-hole symmetry relation Eq.~\eqref{eq:LL_PHsymm}. The former is stable irrespective of $B$ and $\mu$ and given by the WSM system in the first place. Such topology fully determines the reference state $\bar{G}_H$ in the above i) and breaks $\mathcal{T}$ even without $B$.
On the other hand, we note from Eq.~\eqref{eq:LL_PHsymm} that for any empty/filled $\varepsilon_n(B,\mu)\gtrless0$ LL state the corresponding $\varepsilon_{-n}(-B,-\mu)\lessgtr0$ LL is filled/empty, summarized in Table~\ref{table2}. Let's call LL$(B,\mu)$ and LL$(-B,-\mu)$ respectively system I and II. According to Table~\ref{table1}, the reference state has all LL$_{n\leq n_0^b}$ (LL$_{n\leq n_0^{-b}}$) filled in system I (II).
Inspecting the deviation from the reference state due to tunable $\mu\neq0$ according to Table~\ref{table2}, we then realize the following property: when $\mu$ moves up to fill more LLs in system I, any thus filled LL$_{n>n_0^b}$, i.e., $\varepsilon_n(B,\mu)<0$, in system I is paired with an emptied LL$_{-n\leq n_0^{-b}}$, i.e., $\varepsilon_{-n}(-B,-\mu)>0$, in system II; similarly when $\mu$ moves down to fill fewer LLs in system I, any emptied LL$_{n\leq n_0^b}$ in system I is paired with a filled LL$_{-n> n_0^{-b}}$ in system II. Therefore, we maintain a perfect match between the additionally filled (emptied) LLs in system I and those additionally emptied (filled) in system II, which always contribute the \textit{same} conductance quanta due to opposite $B$.
From this analysis, we also know that when the system breaks $\mathcal{P}$ in the first place by any generic term $D$ similar to $\mu$, Eq.~\eqref{eq:symm} will become $G_H(B,\mu,D)=G_H(-B,-\mu,-D)$.

\subsection{Onsager reciprocal relation}
Two additional symmetries can be together cast in a compact form
\begin{equation}\label{eq:symm2}
    G_H(C ,B,\mu)=-G_H(-C ,-B,\mu)
\end{equation}
where $C$ stands for either $t_0,t$ together \textit{or} $\chi$ only. The proof is similar to Eq.~\eqref{eq:symm} as it can be directly verified in the analytic form of Eq.~\eqref{eq:G_H} together with two extra symmetries held by the LLs Eq.~\eqref{eq:LLs2} 
\begin{equation}\label{eq:LLsym1}
    \varepsilon_n(m,t,b)=\varepsilon_n(-m,-t,-b)
\end{equation}
and 
\begin{equation}\label{eq:LLsym2}
    \varepsilon_n(\chi,b)=\varepsilon_n(-\chi,-b).
\end{equation}

The physical meaning of Eq.~\eqref{eq:symm2} is most easily understood collectively as the \textit{Onsager reciprocal relation} respectively for real spins or pseudospins, because the time-reversal operation 
\begin{equation}
    \cT\cH(C,B)\cT^{-1}=\cH(-C,-B)
\end{equation}
respectively for $\cT=\ii\sigma_y K$ or $\cT=K$, where always $\cT\bpi(B)\cT^{-1}=-\bpi(-B)$ due to $K$. 
In this sense, Eq.~\eqref{eq:LLsym1} and Eq.~\eqref{eq:LLsym2} are actually a degeneracy between the $\cT$-related states.
Since our model does not assume the representation basis of the Pauli matrices, we can take advantage that these two cases should both hold. More intuitively and concretely, we can understand it from two things combined: i) $C$ controls the sign of the AHE part Eq.~\eqref{eq:barsigmaHchi} of the reference state while $B$ for the sign of the purely field-induced part Eq.~\eqref{eq:DeltasigmaH}; ii) Mapping $C,B$ to $-C,-B$ does not change the arrangement of the reference state as per the general Table~\ref{table1}.

Certainly, one can in principle stick to one particular representation basis, e.g., the real spin case. The drawback lies in that the symmetry operation $\Gamma=K$, which gives $\Gamma\cH(\chi,B)\Gamma^{-1}=\cH(-\chi,-B)$, same as the pseudospin $\cT$ related to Eq.~\eqref{eq:symm2} when $C$ is $\chi$, is not an obvious one. In fact, $\Gamma=\ii\sigma_y\cT$, i.e., a $\pi/2$ pure spin rotation along $\hat{y}$ combined with the present real-spin $\cT$.

\section{Conventional LL spacing effect and chiral LL mechanism}\label{App:LLspacing}
\subsection{Conventional monotonic LL spacing effect}
Here, we clarify the more conventional effect of varying magnetic field $B$ and the LL spacings. We have already seen such effects in Fig.~\ref{Fig:main} either in the presence or in the absence of the nonmonotonic effect. 

Very large $|B|$ will certainly enlarge $\cm_n$ to invalidate Eq.~\eqref{eq:reshuffle_condition_main} or Eq.~\eqref{eq:reshuffle_condition} and remove nonmonotonic $G_H(B)$ behavior. Intuitively, LL spacings become large and LLs are pushed away from LL$_0$ such that the system eventually returns to the reference state for a given $\mu$.
On the other hand, decreasing $|B|$ reduces LL spacings and LL$_{n\gtrless0}$ are in gross pushed towards LL$_0$ from the above or below respectively, more and more filled or emptied LLs for $\mu\gtrless \bar\varepsilon$ are contributing to $\Delta G_H$ as per Eq.~\eqref{eq:Deltasigma_rule}, leading to decreasing or increasing $G_H$ respectively when 
\begin{equation}\label{eq:LLshrink0}
    b\sgn(\mu-\bar\varepsilon)=\pm.
\end{equation}
Since $\varepsilon_{n\gtrless0}(B\rightarrow0)\sim \mycomment{\bar\varepsilon}\pm|m|$, an alternative viewpoint is that for a fixed $\mu$ those momentum slices with $|\mu|>|m(k_z)|$ will contribute significantly. 

Since $\lim_{|B|\rightarrow0}\bar\varepsilon=0$, care should be taken for the $\mu=0$ case. Actually, given $\sgn(t\chi b)=\sgn(\bar\varepsilon)$, we notice that Eq.~\eqref{eq:LLshrink0} allows $\mu=0$ when $\chi t<0$. Since Eq.~\eqref{eq:LLshrink0} is a necessary condition, the issue is whether such a $\mu=0$ case really accommodates the foregoing strong decreasing or increasing $G_H$. Physically, since $G_H(B\rightarrow0,\mu=0)$ approaches the reference state and returns to the original WSM, one should \textit{not} expect such behavior or any divergence as shown in Fig.~\ref{Fig:other} (hence we exclude $\mu=0$ in Eq.~\eqref{eq:Gdivergence} below). To see this explicitly, we follow the $m$-extremal construction Eq.~\eqref{eq:m_extremal} and thus have the LL extremal energy $\varepsilon_{n\pm}(\mathsf{m},B)=\bar\varepsilon\pm \sqrt{n}v_{B} $ per Eq.~\eqref{eq:LLs} for $n>0$. For small $|B|$, one has $|\bar\varepsilon|<v_{B} $ asymptotically; hence, one always finds that $\varepsilon_{n\pm}(\mathsf{m},B)\gtrless0$ when $\chi t<0$, i.e., by no means one can reduce $|B|$ to let any LL$_{n\neq0}$ cross $\mu=0$ and contribute. Therefore, it is justified to exclude $\mu=0$ case and reduce Eq.~\eqref{eq:LLshrink0}: with decreasing $|B|$, the shrinking LL spacing effect is decreasing (increasing) $G_H$ respectively when 
\begin{equation}\label{eq:LLshrink1}
    b\sgn(\mu)=\pm.
\end{equation}
Such a conventional effect due to reducing LL spacing is almost always present. For instance, it contributes to the overall trend in Fig.~\ref{Fig:main}(c,d) and becomes very significant when $B\rightarrow0$ as expected. On the other hand, when the nonmonotonic condition Eq.~\eqref{eq:NMLL_0_condition} is not satisfied, the nonmonotonic Hall effect is missing as predicted and substituted by such conventional monotonic behavior, as shown in Figs.~\ref{Fig:other}(a,b) for the $b>0,\mu\leq0$ and $b<0,\mu\geq0$ cases, respectively. Clearly shown in these data, the $\mu=0$ case in general does not bear the strong variation approaching $B=0$ and merely returns to the original WSM AHE, just as we found above.

\begin{figure}[t]
\includegraphics[width=17.8cm]{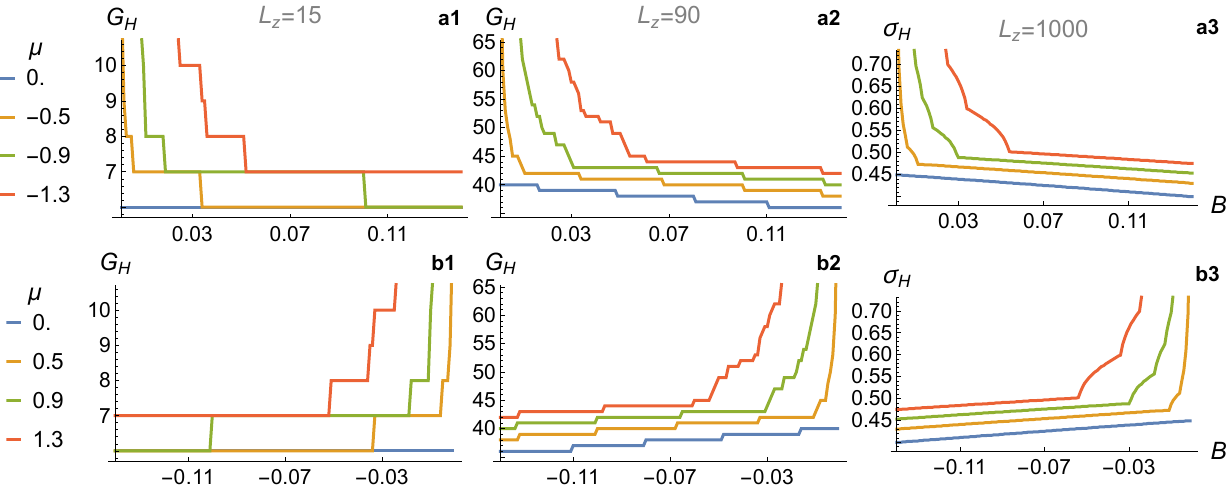}
\caption{(a,b) Monotonic Hall signals when $b\mu\leq0$, in stark contrast to nonmonotonic Figs.~\ref{Fig:main}(c,d), dependent on chemical potential $\mu$ and thickness $L_z$: (a1-2,b1-2) thin films or mesoscopic samples (conductance $G_H$ in units of $e^2/\hbar$); (a3,b3) bulk samples (3D conductivity $\sigma_H=G_H/L_z$).
Parameters are the same as those of Figs.~\ref{Fig:main}(c,d). 
Either (a1-3) $b>0,\mu\leq0$ or (b1-3) $b<0,\mu\geq0$ only exhibits the conventional monotonic Hall effect due to the LL spacing reduction, without NM~I and II at work. Comparison between (a) and (b) also exemplifies the symmetry relation Eq.~\eqref{eq:symm_main}.
Note that the $\mu=0$ case neither shows nonmonotonicity nor diverges approaching $B=0$.
}\label{Fig:other}
\end{figure}

In fact, since $\lim_{|B|\rightarrow0}\bar\varepsilon=0$, the conductance will diverge as 
\begin{equation}\label{eq:Gdivergence}
    G_H(B\rightarrow0,\mu\neq0)\sim-b\sgn(\mu)/|B|,
\end{equation}
because the number of contributing LLs $\sim1/\hbar\omega_c=|\varepsilon_{n+1}-\varepsilon_{n}|^{-1}\sim (v_{B} ^2/2m)^{-1}\propto|B|^{-1}$ for small $|B|$ as per Eq.~\eqref{eq:LLs2}, where we define the cyclotron frequency $\omega_c$ from the LL spacing.
%Note that $G_H(B\rightarrow0,\mu=0)$ does not diverge as it approaches the reference state.
In reality, although this will be eventually regularized to a Drude form for each contributing momentum slice as $ G_H\sim|B|\tau$ by scattering mechanisms giving rise to a relaxation time $\tau$, the drastic decreasing or increasing behavior of $G_H$ will remain.
To see this regularized form, we replace $\frac{n}{m}$ in the dc Drude formula $G_0=\frac{n}{m}e^2\tau$ for quadratic band electrons by $\frac{\omega_c}{\hbar}$, because for the Hall effect of such electrons one has $\omega_c=\frac{eB}{m}=\frac{\hbar}{l_{B}^2m}=\frac{n\hbar}{m}$. Then, substituted by the $\omega_c$ defined above for our case, it leads to $ G_H=\omega_c\tau\frac{e^2}{\hbar}$ up to the leading order expansion in terms of $\omega_c\tau$. Hence, given $\tau$, $ G_H(B\rightarrow0)\sim|B|$ decreases linearly apart from the AHE. Actually, these consideration becomes pertinent when $\omega_c\tau<1$, i.e., $|B|<|B|_c\sim\frac{\hbar^2m_0}{e\tau t_1^2}$. 
On the one hand, such an inevitable peak structure or monotonic behavior approaching small $|B|$ can serve as one cause of nonmonotonic $G_H(B)$ when $|B|$ decreases as long as the behavior at larger $|B|$ differs. On the other hand, it may eventually wash out the quantum nonmonotonicity because effectively $|\mu|$ becomes large enough to be far away from the bent/reshuffled LLs as aforementioned.

\subsection{Nonmonotonicity due to chiral LL effect}
Now we can find the generic chiral LL$_0$ mechanism for nonmonotonicity.
Seen from Eq.~\eqref{eq:DepsilonDB}, $\frac{\partial\varepsilon_0}{\partial|B|}=e t\chi b/2\hbar$, which shifts upward/downward linearly as a whole with decreasing $|B|$ respectively for $\sgn(t\chi b)=\mp$. Correspondingly, as per Eq.~\eqref{eq:Deltasigma_rule}, for a given quantized $m(k_z)$ slice and decreasing $|B|$, $\Delta G_H$ will change by $-\sgn(t\chi)\frac{e^2}{h}$ once LL$_0$ moves across $\mu$, \textit{regardless} of $b$ or the moving direction. 
Given Eq.~\eqref{eq:LLshrink1}, this chiral LL$_0$ effect of changing $G_H$ is constructive (destructive) to the conventional shrinking LL spacing effect of varying $G_H$ respectively when $\sgn(t\chi b\mu)=\pm$. 
Therefore, only in the case of 
\begin{equation}\label{eq:NMLL_0_condition_generic}
    t\chi b\mu<0
\end{equation}
can nonmonotonic $G_H(B)$ appear. 
It is the counterpart of Eq.~\eqref{eq:NMLL_0_condition}.
This condition encompasses, i.e., is a \textit{necessary} condition of, the rightmost part of Eq.~\eqref{eq:m&mu_condition_general} as expected.
Therefore, it is also the most general necessary condition of nonmonotonic behavior taking \textit{all} mechanisms into account. 

This NM~II will also modify SdHE.  This is because the latter is due to the conventional LL spacing variation and thus follows the monotonic behavior. Considering decreasing $|B|$,
the SdHE cusps will in general be delayed to fields smaller than $|B|_n$ predicted by pure SdHE effects.
As LL$_0$ lies close to the first few LLs, it will affect the initial SdHE more at larger fields, e.g., the first SdHE cusp deviates from $|B|_1$ the most. Exemplified by the $\mu=0.6$ green curve in Fig.~\ref{Fig:main}(d3) with decreasing $|B|$, before the rightmost cusp it shows the anomalously increasing Hall signal; later at smaller $|B|$ typically after this cusp, the monotonic SdHE prevails and they together exhibit nonmonotonicity. Due to the chemical potential dependence that smaller $|\mu|$ asks for smaller $|B|$ to touch a new LL, such anomalous region delaying the monotonic SdHE becomes wider at smaller $|\mu|$ as Fig.~\ref{Fig:main}(d3) shows.

\section{LL bending and reshuffling mechanism}\label{App:LLbending}

\subsection{Derivation of LL bending and LL energy surface gradient}\label{App:LLbending_cond}
Now we derive the quantitative condition of LL bending for $n\neq0$. We consider $m>0$ only as the LLs are otherwise obviously monotonic without bending. As the mechanism suggests, a bending LL crosses $\mu$ twice along $|B|$-axis, which requires that 
\begin{equation}\label{eq:mu=epsilon}
    \mu=\varepsilon_n(\chi b, u_{B})
\end{equation}
has two positive solutions as an equation of $|B|$. The solutions are 
\begin{equation}
    |B|_\pm=\frac{2\hbar}{et}\frac{\gamma_n\pm\sqrt{\Delta^{B}_n}}{4n^2-1}
\end{equation}
with $\gamma_n=2|n|(m-\cm_0)-\chi b\mu$
and $\Delta^{B}_n=\gamma_n^2-4(4n^2-1)(m^2-\mu^2)$.
The existence of $|B|_\pm$ requires $\Delta^{B}_n>0$. In order to show systematic nonmonotonic behavior, it is necessary to let as many $k_z$'s, i.e., different $m$'s, contribute such nontrivial bending effect as possible. Thus, we require a sufficient condition such that 
\begin{equation}
    \Delta^{B}_n(m)>0
\end{equation}
holds for any $m$. This leads to its determinant $\Delta^m_n=t_1^2|n|(4n^2-1)(\chi b\mu t+|n|t_1^2)<0$, i.e., %$b\mu<-|n|\cm_0$.
\begin{equation}\label{eq:B_exist}
    \sgn(t)\chi b\mu<-\sgn(t)|n|\cm_0.
\end{equation}
The positivity of $|B|_\pm$ requires $\sgn(t)\gamma_n>0$ and $\Delta^{B}_n<\gamma_n^2$, leading to 
\begin{equation}\label{eq:B_positivity}
    \sgn(t)m>\sgn(t)\left(\cm_0+\frac{\chi b\mu}{2|n|}\right),\quad m>|\mu|.
\end{equation}
Combining them all together, we find a concise condition to have nonmonotonicity in $G_H(B)$ 
\begin{equation}\label{eq:m&mu_condition_general}
    \sgn(t)m_0>\sgn(t)m>-\sgn(t)\chi b\mu \gtrsim \sgn(t)\cM_n.
\end{equation}
Notice that we assume $\sgn(t)=\sgn(t_0)$ for the leftmost inequality, otherwise the system is not even a WSM and becomes trivial. 
Once the magnetic ordering is reversed, e.g., by external magnetic fields, the system will have $t_0,t<0$.
We use $\gtrsim$ to indicate that the last inequality is a sufficient condition such that a slight deviation is also possible. 
This reduces to Eq.~\eqref{eq:m&mu_condition} when $t>0$, where we do not indicate $\gtrsim$ explicitly for brevity.
For example, given $\mu$ and $n$, the more $m(k_z)$'s of allowed momentum slices satisfy Eq.~\eqref{eq:m&mu_condition} due to higher $m_0$, the more contribution to the overall nonmonotonicity. On the other hand, given $m_0$, although higher $|\mu|$ and hence higher $|n|$ implies more bending LLs involved, the nonmonotonicity of $G_H(B)$ will initially increase with $|\mu|$ but eventually be suppressed due to less and less satisfying $m(k_z)$'s.

We also clarify the meaning of LL bending with the slope or gradient of the LL energy surface with respect to the magnetic field 
\begin{equation}\label{eq:DepsilonDB}
    \frac{\partial\varepsilon_n}{\partial|B|}=\frac{e t}{\hbar}\left[\frac{\chi b}{2}+\frac{n(\cm_{2n}-m)}{\kappa_{|n|}}\right]
\end{equation}
where
\begin{equation}\label{eq:tildem_n}
    \cm_{n}(|B|)%=\frac{t_1^2}{t}+|n| u_{B}
    =\frac{t_1^2}{t}+\frac{|n|t}{2l_{B}^2}.
\end{equation} 
Then the LL bending means the sign change of Eq.~\eqref{eq:DepsilonDB} as $|B|$ is varied. 
In the main text, we mention that such bending does not happen for several trivial cases: $t=0$ and small $m$ or large $|B|$ such that Eq.~\eqref{eq:reshuffle_condition} is violated.
To see them, one notes from Eq.~\eqref{eq:DepsilonDB} that the corresponding LL slopes 
\begin{equation}
    \frac{\partial\varepsilon_n}{\partial|B|}|_{t=0}=\frac{e t_1^2n}{\hbar\sqrt{m^2+|n|v_{B} ^2}},
\end{equation}
\begin{equation}
    \frac{\partial\varepsilon_n}{\partial|B|}|_{m\rightarrow0}=\frac{et}{\hbar}(\frac{\chi b}{2}+\frac{n\cm_{2n}}{\sqrt{\cm_{2n}^2-\cm_0^2}}),
\end{equation}
and 
\begin{equation}
    \frac{\partial\varepsilon_n}{\partial|B|}|_{|B|\rightarrow\infty}=\frac{et}{\hbar}(\frac{\chi b}{2}+n)
\end{equation}
share the same fixed sign as $n$ except for LL$_0$, i.e., there is no LL bending effect contributing to NM~I in these cases.
Lastly, we note that looking at the slope alone does not take the chemical potential $\mu$ into account.

\subsection{Relation between LL bending and reshuffling}\label{App:LLslope}
It is interesting to further understand the effect of $m$ being larger than its lower bound $\cM_1$ in Eq.~\eqref{eq:m&mu_condition_general}; a drawback thereof is the concealed $|B|$-dependence at the price of explicitly incorporating the chemical potential.
Taking a complementary view, it connects to the reshuffling of LL ordering. 
Next, we derive the condition for 
the phenomenon of LL reshuffling $\varepsilon_{n+1}<\varepsilon_n$ for general $n\in\mathbb{N}$, i.e., it violates the normal LL ordering $\varepsilon_{n+1}>\varepsilon_n$. Such reshuffling is a natural consequence of the bending LLs, because they can cross each other as $|B|$ or $m(k_z)$ varies and hence change their relative ordering. 

Firstly, we find it never fulfilled in the $tm<0$ case however $b$ is. Secondly and more interestingly for the $tm>0$ side, there are four cases. i) $n>0$ gives $\sgn(t)m>\sgn(t)[\frac{t_1^2}{t}+(2n+1) u_{B}]$; ii) $n=0$: $\sgn(t\chi b)=+$ not possible since $\partial_m\varepsilon_0(m)=-\chi b$ gives $-\sgn(t)$ chirality of LL$_0$ while $\sgn(t\chi b)=-$ gives $\sgn(t)m>\sgn(t)[\frac{t_1^2}{t}+ u_{B}]$; iii) $n=-1$: $\sgn(t\chi b)=-$ not possible since $\partial_m\varepsilon_0(m)=-\chi b$ gives $\sgn(t)$ chirality of LL$_0$ while $\sgn(t\chi b)=+$ gives $\sgn(t)m>\sgn(t)[\frac{t_1^2}{t}+ u_{B}$]; iv) $n<-1$ gives $\sgn(t)m>\sgn(t)[\frac{t_1^2}{t}-(2n+1) u_{B}]$. These combined, we obtain  
\begin{equation}
\sgn(t)m>\sgn(t)[\frac{t_1^2}{t}+\frac{|2n+1|t}{2l_{B}^2}]
\end{equation}
and hence 
\begin{equation}\label{eq:reshuffle_condition}
    \sgn(t)m_0 > \sgn(t)m>\sgn(t)\cm_{2n+1}(|B|),
\end{equation}
which acquires a $|B|$-dependent mass $\cm_{n}(|B|)%=\frac{t_1^2}{t}+|n| u_{B}
    =\frac{t_1^2}{t}+\frac{|n|t}{2l_{B}^2}$.
It reduces when $t>0$ to
\begin{equation}\label{eq:reshuffle_condition_main}
    m_0 > m(k_z)>\cm_{2n+1}(|B|).
\end{equation}
% with a threshold mass dependent on the LL index
% \begin{equation}\label{eq:tildem_n}
%     \cm_n = \frac{v_{B} ^2}{4 u_{B}}+|2n+1| u_{B} = \frac{t_1^2}{t} +\frac{e\hbar t}{2}|2n+1||B|.
% \end{equation}
Note that when $n=0,-1$, its validity respectively requires $\sgn(t\chi b)=\mp$. Since $\cm_n$ %given in Eq.~\eqref{eq:tildem_n} 
increases with $|n|$, the larger $|n|$ Eq.~\eqref{eq:reshuffle_condition_main} satisfies, the more LL reshuffling can happen. 

Also suggested by Eq.~\eqref{eq:reshuffle_condition_main}, the trivial ordering $\varepsilon_{n+1}>\varepsilon_n$ is seen at small $m$ or large $|B|$ and simply follows the LL index, which does not contribute to the nonmonotonicity in $G_H(B)$. %(\ref{App:LLslope}). 
To fully appreciate the dependence on $t$ together with the magnetic field, we note the following properties. Consistent with the extreme case, too small $t$ will always make $\cm_n$ large enough to never fulfill Eq.~\eqref{eq:m&mu_condition} and Eq.~\eqref{eq:reshuffle_condition_main}. In such a case, although generically we do not expect nonmonotonic $G_H(B)$ from LL bending and reshuffling, there remains the other chiral LL$_0$ mechanism of nonmonotonicity. However, Eq.~\eqref{eq:m&mu_condition} does not forbid large $t$ because it can always be compensated by small $|B|$ to still maintain moderate $\cm_n$ in Eq.~\eqref{eq:reshuffle_condition_main} and nonmonotonic $G_H(B)$ can still occur. 

To mathematically see the relation between LL bending and LL reshuffling, we rely on Eq.~\eqref{eq:DepsilonDB}. For simplicity, we assume $t>0$. As aforementioned, at large magnetic field $|B|$, the LL slope becomes $\frac{\partial\varepsilon_n}{\partial|B|}|_{|B|\rightarrow\infty}=\frac{e t}{\hbar}(\frac{\chi b}{2}+n)$ and its sign is the same as $\sgn(n)$ for all LLs except the special LL$_0$. Therefore, LL bending can also be formulated as 
\begin{equation}
    \frac{\partial\varepsilon_n}{\partial|B|}\sgn(n)<0,
\end{equation}
which leads to 
\begin{equation}\label{eq:DepsilonDB_condition}
    m-\cm_{2n}>\chi b\frac{\kappa_{|n|}}{2n}.
\end{equation}
For $\chi b\sgn(n)<0$, Eq.~\eqref{eq:DepsilonDB_condition} is systematically satisfied as long as $m-\cm_{2n}>0$, which is identical to the LL reshuffling condition Eq.~\eqref{eq:reshuffle_condition} up to a difference in the LL index. This is physically clearly corresponding to the $\chi B\mu<0$ condition we found from Eq.~\eqref{eq:m&mu_condition}, because the majority of LL$_{n\neq0}$ have $\varepsilon_{n\gtrless0}\gtrless0$ especially for small magnetic fields.
On the other hand, for $\chi b\sgn(n)>0$, it gives 
\begin{equation}
    m>2|n| u_{B}+\frac{1}{4n^2-1}\left(  4n^2\cm_0 +\sqrt{4n^2\cm_0^2+|n|(4n^2-1)v_{B} ^2} \right).
\end{equation}
Such a condition of $m$ is obviously not as easily and systematically satisfied as the foregoing case and imposes a much more stringent constraint on the contributing $m(k_z)$ slices. Similarly, this roughly corresponds to the $\chi B\mu>0$ case where nonmonotonicity is largely accidental and subject to very limited parameter ranges.

\subsection{Nonmonotonicity under more limited conditions}\label{App:exceptions}
Beyond the two symmetry-related cases of $\chi B\mu<0$ for a given $\chi$, there remain some much rarer situations where nonmonotonic $G_H(B)$ also can occasionally appear when $\chi B\mu>0$ or $\mu=0$.
They are both related to the accurate solution of $\Delta^{u_{B}}_n(m)>0$ for Eq.~\eqref{eq:mu=epsilon} for each \textit{specific} $m$: $\mu\gtrless\mu_\pm(m)$ with 
\begin{equation}
    \mu_\pm(m)=\frac{\chi bt(m-\cm_0) \pm t_1\sqrt{t(4n^2-1)(2m-\cm_0)}}{2|n|t}.
\end{equation}
Therefore, they also belong to the LL bending situation.

The first case is for general parameter ranges when $\chi B\mu>0$, where only accidental and tiny bumps can very rarely appear. The reason is for a given $\mu$ the quantized $m(k_z)$ slices satisfying such a condition are very limited and easily buried in the momentum summation; also, they will be strongly compensated by the conventional monotonic contributions from other $m$'s that are of much less constraint. Hence, such nonmonotonic bumps will be entirely washed out by the widely present conventional monotonic contributions in the thick sample case. In stark contrast, the mechanism in the main text under the condition $\chi B\mu<0$ provides systematic nonmonotonic contributions from \textit{all} quantized $m(k_z)$ slices as long as other conditions are met. This leads to a qualitative difference in the end.

The second case is not accidental but only appears in extreme parameter ranges that are often unrealistic. Taking the $\chi B>0$ case as an example, the range of chemical potential $0\leq\mu<\chi b u_{B}$ goes beyond the $\chi B>0,\mu<0$ constraint we have focused on. However, it can possibly exhibit nonmonotonic $G_H(B)$ when $t$ is very large compared to other parameters and the contributing LLs are typically with indices close to $-1$ or LL$_{-1}$ only. This actually corresponds to the above $\mu<\mu_-(m)$ condition when $\mu_->0$. This condition of $\mu_-$ leads to $m-\cm_0>0$ and 
\begin{equation}\label{eq:cond_exception}
    \frac{(m-\cm_0)^2}{2m-\cm_0}>(4n^2-1)\cm_0.
\end{equation}
From this, we immediately see that small $|n|$ and large enough $t$ will accommodate more different $m(k_z)$ slices since $\cm_0=\frac{t_1^2}{t}$.
As a special example, it can support nonmonotonic $G_H(\chi B>0,\mu=0)$ at the original WP energy with large enough $t$, which is mainly caused by the $n=0,-1$ cases of Eq.~\eqref{eq:reshuffle_condition}.
Also, note that this mechanism survives in thick samples because it provides a less constrained choice of quantized $m(k_z)$ slices in a similar manner to the earlier situations we discussed.

\section{Detailed formulation and proof related to SdHE}\label{App:SdHE}

\subsection{SdHE due to pure LL spacing variation}\label{App:SdH_cusp}
Here we formulate in detail the effect of LL spacing variation on SdHE. 
The cusp point $B_n$ due to LL$_{n\neq0}$ is determined by
\begin{equation}\label{eq:osci1}
    \sgn(n)\mu=\min_{m\leq m_0} \sgn(n)\varepsilon_{n}(m,B_n),
\end{equation} 
because the corresponding cusp- or kink-like change in $G_H(B)$ will appear once $\mu$ begins to cross this LL in \textit{any} $m(k_z)$ slice, since every LL$_{n\neq0}$ as a function of $m$ has one extremum subject to the mass bound $m_0$. 
Based on Eq.~\eqref{eq:osci1}, representatively, we can require $m_0$ to accommodate the LL extremum within the bound $m_0$, i.e., at $\mathsf{m}=2|n| u_{B}<m_0$ where 
\begin{equation}\label{eq:m_extremal}
    \frac{\partial\varepsilon_n}{\partial m}=0,
\end{equation}
which is a reasonable and relevant situation as it is also implied by our LL reshuffling condition Eq.~\eqref{eq:reshuffle_condition}. Then Eq.~\eqref{eq:osci1} becomes  
\begin{equation}\label{eq:osci2}
    \mu=\varepsilon_{n}(\mathsf{m},B_n).
\end{equation}
Formally, when $\chi b\mu>0$ the only solution is $B_n=b|B|_{n,-\sgn(\mu n)}$; when $\chi b\mu<0,\mu n>0$ there are two solutions $B_n=b|B|_{n\pm}$ and no solution otherwise. Here, we define 
\begin{equation}\label{eq:Bn_sol}
    %|B|_{n\pm}^{-1}=\frac{e}{\hbar}\frac{2|n|t_1^2+tb\mu \pm 2t_1\sqrt{|n|(|n|t_1^2+tb\mu)}}{2\mu^2}
    |B|_{n\pm}=\frac{2\hbar}{e}\frac{2|n|t_1^2+t\chi b\mu \pm 2t_1\sqrt{|n|(|n|t_1^2+t\chi b\mu)}}{t^2}
\end{equation}
satisfying $|B|_{n+}|B|_{n-}=\frac{4\hbar^2\mu^2}{e^2t^2}$ and $|B|_{n+}>|B|_{n-}>0$ as long as Eq.~\eqref{eq:osci_cusp_exist} guarantees the existence.
From these relations, we know that the smaller $|B|_{n-}$ is physically more relevant and hence usually the experimentally achievable magnetic field while $|B|_{n+}$ is typically unrealistically large. The rare exception might happen, e.g., for small $|\mu|$ such that $|B|_{\pm1,+}$ is not too far away from the experimental observation window, and between $|B|_{\pm1,+}$ and $|B|_{\pm1,-}$ is the initial decreasing before reaching $|B|_{\pm1,-}$ as part of the nonmonotonic turn. 
Note that we have $|B|_{n-}(t=0)=\frac{\hbar\mu^2}{2e|n|t_1^2}$ and $|B|_{n+}(t=0)=\infty$, i.e., only $|B|_{n-}$ is connected to the $t=0$ case with only one solution as expected. Therefore, although finite $t$ gives rise to extra cusp points in the SdHE, often only the one at a smaller field can be observed. Henceforth and in the main text, we will simply refer $|B|_{n-}$ as $|B|_n$.
The cusp field $|B_n|$ based on Eq.~\eqref{eq:osci1} will only be slightly lower than $|B|_{n}$ when $\mathsf{m}$ exceeds the bound $m_0$, because a smaller magnetic field is necessary to push LL$_n$ across $\mu$ at $m_0$.

Now the finite $t$ crucial to WSM situation makes the separation $\Delta |B|_n^{-1}=|B|_{n+\sgn(n)}^{-1}-|B|_n^{-1}$ between neighboring cusps nonuniform, in contrast to the conventional $t=0$ case with a constant $\Delta |B|_{t=0}^{-1}=2et_1^2/\hbar\mu^2$. Such deviation becomes less pronounced for LLs with higher index $|n|$, i.e., those at smaller magnetic fields, which can be seen from $\Delta |B|_{n\rightarrow\infty}^{-1}=\Delta  |B|_{t=0}^{-1}$. 

%\subsection{Sufficient condition of SdHE}\label{App:SdHEsuffcond}
Eq.~\eqref{eq:osci_cusp_exist} is actually a \textit{sufficient} condition for the SdHE due to LL$_n$, 
similar to the last part of Eq.~\eqref{eq:m&mu_condition} for LL$_n$ bending.
Now, we prove that Eq.~\eqref{eq:osci_cusp_exist} serves as a sufficient condition of general SdHE, especially including the case when $m_0$ is not large enough to reach the extremal $\mathsf{m}$.
It is actually the extremum of LL$_{n\neq0}$ that is more demanding for a fixed $\sgn(n)\mu$ to reach by varying $B$ than any higher part of $\sgn(n)\varepsilon_n(B,m)$. The higher bound of the $\sgn(n)$ attached LL$_n$ energy surface is
\begin{equation}
\begin{split}
    E_h=&\max_{m\in[0,\mathsf{m}]}\sgn(n)\varepsilon_n(B,m)\\
    =&\sgn(n)\varepsilon_n(B,0)=\chi b u_{B}\sgn(n)+\sqrt{4n^2 u_{B}^2+|n|v_{B} ^2}>0    
\end{split}
\end{equation}
and can be made arbitrarily close to $0$ from the above by reducing $|B|$; the extremum as its lower bound 
\begin{equation}
    E_l=\sgn(n)\varepsilon_n(B,\mathsf{m})=\chi b u_{B}\sgn(n)+\sqrt{|n|}v_{B} <E_h
\end{equation}
but is \textit{not} necessarily positive. 

For the more complex case $m_0<\mathsf{m}$, the relevant energy $E_l<\sgn(n)\varepsilon_n(B,m_0)<E_h$ lies in between and can always be reached by reducing $|B|$, as long as $E_l$ is positive and can be reached. This is just equivalent to requiring $\mu n>0$ and Eq.~\eqref{eq:osci_cusp_exist}. 
The extra constraint $\mu n>0$ makes explicit the contributing LL between LL$_{\pm n}$ and excludes the special case $\sgn(n)\varepsilon{n}<0$ possible for lower $|n|$; if $\mu n<0$ and Eq.~\eqref{eq:osci_cusp_exist}, the SdHE is limited to a few lower $|n|$ and misses the majority of LLs since $\mu$ is on the `minority' side.

\subsection{Relation between SdHE and LL bending}\label{App:SdHEandLLbending}
Here, we give a rigorous proof for the mutual exclusiveness between the LL$_n$ SdHE condition Eq.~\eqref{eq:osci_cusp_exist} and the LL$_n$ bending condition Eq.~\eqref{eq:m&mu_condition}. The proof contains two steps: firstly we show the equivalence between the SdHE condition and the existence of vertical tangent points in the $\mu$-cut of the LL energy surface in the $(m,|B|,\varepsilon)$-space; secondly, we show that the latter must violate the systematic LL bending requirement that for each $m(k_z)$ slice $\mu$ crosses the LL twice.

\begin{figure}[hbt]
\includegraphics[width=6.8cm]{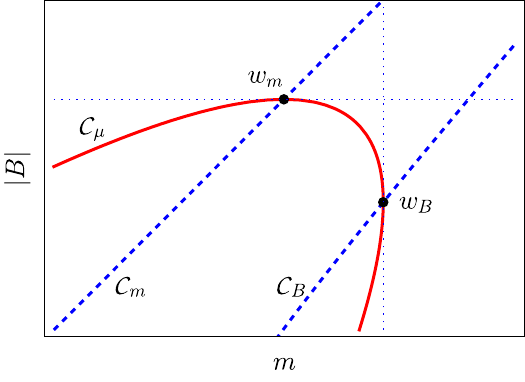}
\caption{Illustration of the representative geometry of extremal curves of an LL energy surface projected onto the $(m,|B|)$-plane. Red solid curve $\cC_\mu$: chemical potential $\mu$-cut of the LL energy surface. Blue dashed curve $\cC_{B}$ ($\cC_m$): $|B|$-extremal ($m$-extremal) curve where LL energy is extremal with respect to $|B|$ ($m$). Black point $w_{B}$ ($w_m$), where $\cC_{B}$ ($\cC_m$) crosses $\cC_\mu$, accommodates a vertical (horizontal) tangent line of $\cC_\mu$, as indicated by the blue dotted line.
}\label{Fig:Extr}
\end{figure}

\textit{Step I}: Similar to the construction of $m$-extremal curves as per Eq.~\eqref{eq:m_extremal}, we also find the $|B|$-extremal curves as shown in Fig.~\ref{Fig:Extr}. The extremum along $|B|$ is given by 
\begin{equation}\label{eq:B-extremal}
    \frac{\partial\varepsilon_n}{\partial |B|}=0,
\end{equation}
leading to \begin{equation}\label{eq:Bextrem-B}
    \mathsf{B}_{\chi bn}=\frac{\hbar}{e}\frac{m-\cm_0-\chi b\sgn(n)\sqrt{\frac{\cm_0}{4n^2-1}(2m-\cm_0)}}{t|n|}
\end{equation} 
with $\mathsf{B}$ denoting $|B|$ at the extremum.
We have $\mathsf{B}_{\chi bn>0}>0$ when $m>\cm_0$ and $\mathsf{B}_{\chi bn<0}>0$ when $m>2\cm_0$ for any LL index $n$, which is a very weak and reasonable requirement as readily implied by even the minimal LL reshuffling condition Eq.~\eqref{eq:reshuffle_condition} evaluated at $|n|=1$. Then we require the chemical potential to cross the corresponding LL extremum
\begin{equation}
    \mu=\varepsilon_n(\mathsf{B}_{\chi bn})=\frac{\chi b\sgn(n)(m-\cm_0)+\sqrt{\cm_0(4n^2-1)(2m-\cm_0)}}{2n},
\end{equation}
whose solution 
\begin{equation}\label{eq:Bextrem-m}
    m_\pm=2[2n^2\cm_0+\chi b\mu|n| \pm \sqrt{|n|(4n^2-1)\cm_0(\cM_n+\chi b\mu)}]
\end{equation}
exists under the very \textit{same} condition as Eq.~\eqref{eq:osci_cusp_exist} and also satisfies always $m_\pm>0$ as expected. This completes the first step of the proof.

\textit{Step II}: We consider three curves in the $(m,|B|,\varepsilon)$-space projected onto the $(m,|B|)$-plane as shown in Fig.~\ref{Fig:Extr}. The first $\cC_\mu$ is the crossing curve between the horizontal plane $\varepsilon=\mu$ and the LL$_n$ energy surface $\varepsilon=\varepsilon_n(m,B)$. The other two curves $\cC_{B},\cC_m$ are respectively the $|B|$-extremal curve $\varepsilon=\varepsilon_n(m,b\mathsf{B}(m))$ and the $m$-extremal curve $\varepsilon=\varepsilon_n(\mathsf{m}(B),B)$ on the energy surface and projected onto the $(m,|B|)$-plane, which extend %from the origin $(m=0,B=0)$ 
towards the infinity of both $m$ and $B$. Eq.~\eqref{eq:osci_cusp_exist} is then equivalent to the existence of at least one crossing at some $w_{B}=(m_{B},|B|_{B})$ and $w_m=(m_m,|B|_m)$ respectively between $\cC_\mu$ and $\cC_{B}$ and between $\cC_\mu$ and $\cC_m$. Importantly, since $\frac{\partial\varepsilon_n}{\partial |B|}|_{w_{B}}=0$ as per $w_{B}\in\cC_{B}$ and $\frac{\partial\varepsilon_n}{\partial m}|_{w_m}=0$ as per $w_m\in\cC_m$, the vertical (horizontal) line $m=m_{B}$ ($|B|=|B|_m$) becomes a \textit{tangent} line of $\cC_\mu$ on the $\varepsilon=\mu$ plane.

Now we make use of this vertical tangent $m=m_{B}$ of $\cC_\mu$ on the $\varepsilon=\mu$ plane, i.e., the vertical dotted line in Fig.~\ref{Fig:Extr}. On one neighborhood side of this tangent line, e.g., its left side, if the vertical mass line $m=m_{B}- \delta m<m_{B}$ cuts $\cC_\mu$ twice, the other mass line $m=m_{B}+ \delta m>m_{B}$ on the right must not cut $\cC_\mu$. On the other hand, if $w_{B}$ happens to be an inflection point of $\cC_\mu$, typically on both neighborhood sides the mass lines cut $\cC_\mu$ only once; otherwise, there must be another non-inflection $w_{B}'$ such that the slope can change its sign and hence makes more than one cuts possible, which then goes back to the former case. Therefore, we prove that there always exists a mass line not cutting $\cC_\mu$ twice, i.e., breaking the LL bending condition; this completes the second step of the proof and rigorously explains the exclusiveness between Eq.~\eqref{eq:m&mu_condition} and Eq.~\eqref{eq:osci_cusp_exist}.

\textit{Discussion}: The above proof formally only relies on the existence of $w_{B}$. To view the situation more intuitively, we can make full use of the fact that $\cC_\mu$ possesses both a vertical tangent point $w_{B}$ and a horizontal tangent point $w_m$ as aforementioned and illustrated in Fig.~\ref{Fig:Extr}. For a smooth curve $\cC_\mu$, let us consider the typical case without an inflection point and then the tangent $\tan\theta$ of $\cC_\mu$ must change its sign twice continuously, e.g., rotating clockwise from $\theta>0$ to $w_m$ ($\theta=0$), $w_{B}$ ($\theta=-\frac{\pi}{2}$), and finally $\theta<-\frac{\pi}{2}$. In such a situation, $\cC_\mu$ must cross both the $|B|$-axis and the $m$-axis and it obviously leaves an `open' region of $m$ that is beyond $\cC_\mu$, where $\cC_\mu$ is never crossed by $m$-cuts and thus violates the LL bending requirement. Note that this open region may be either semi-infinite or finite in $m$. The former gives an arc-like $\cC_\mu$ and thus has only one $w_m$ or $w_{B}$; the latter $\cC_\mu$ comprises two disconnected arcs and has two sets of $w_m,w_{B}$. In terms of $w_m$, they simply respectively correspond to the foregoing two cases of Eq.~\eqref{eq:osci2} with one or two formal solutions of $|B|_n$.

We additionally can comment on another possible question: instead of the $m$-extremal viewpoint, can one adopt the $|B|$-extremal approach used in this proof to discuss the SdHE and the cusp positions? The answer is negative. 
Firstly, for the SdHE due to varying magnetic field, the physical approach is to find the LL extremum at fixed $B$ and tune $B$ to let it be reached by the chemical potential $\mu$, i.e., the $m$-extremal viewpoint. The $|B|$-extremal approach does not correctly capture this SdHE process, it instead describes a \textit{distinct} SdHE due to varying $m$ as given by Eq.~\eqref{eq:Bextrem-m}, which is not experimentally relevant. Secondly, the foregoing horizontal $w_m$ and vertical $w_{B}$ tangent points obviously bear distinct geometric meanings and are different; this can also be seen from the mismatch between Eq.~\eqref{eq:Bn_sol} and the combination of Eq.~\eqref{eq:Bextrem-B} and Eq.~\eqref{eq:Bextrem-m}, i.e., the $|B|$-extremal result.

Before closing this section, we briefly exemplify the nonexclusiveness of the two phenomena, SdHE and NM~I,  explained in the main text, using the green curve in Fig.~\ref{Fig:main}(c3). This case corresponds to $\mu=1.8$ and we readily find $\cM_2<-\chi b\mu<\cM_3$. According to Table~\ref{tab:NM_SdH_relation}, this can accommodate SdHE due to LL$_{l\geq3}$ and simultaneously NM~I due to LL$_{l<3}$. Indeed, the nonmonotonic turns in the green curve in Fig.~\ref{Fig:main}(c3) are due to LL$_1$ and LL$_2$. On the other hand, the circled SdHE cusps from right to left are successively due to LL$_{3,4,\dots}$. Note that LL$_{1,2}$ does not lead to physical solutions for SdHE as per Eq.~\eqref{eq:Bn_sol}. Therefore, this case does show the coexistence between NM~I and SdHE.

\section{Experimental considerations}\label{App:Experiment}
\subsection{Parameters and scalings}\label{App:Parameters}
While our formulae are general, we can also assume the lattice constant $a=1$ for simplicity in the calculation, i.e., measuring any quantity with a length dimension in the unit of the lattice constant. This includes the parameters $t_1,t,t_0$ in Eq.~\eqref{eq:h} and the sample thickness $L_z$ in Eq.~\eqref{eq:k_i} and Eq.~\eqref{eq:i*} according to their dimensions and the magnetic length $l_{B}$, which is the only combination of $e,\hbar,|B|$ that appears in our theory.
This way, any $t_1,t,t_0$ values we put in the numerical evaluation are just $\bar t_1,\bar t,\bar t_0$, where we use a bar to explicitly mark that they are for Eq.~\eqref{eq:H}.
If we instead want to change from $a=1$ to some other value, e.g., using the real lattice constant, we need to put explicit length scaling factors so as to keep the result \textit{invariant}. This is done by replacing $t_1,t,t_0,l_{B},L_z$ by $t_1a,ta^2,t_0a^2,l_{B} a,L_za$, with which one can mathematically confirm the scaling invariance of all the main results. Since we often put $e=\hbar=1$ and use $B$ instead of $l_{B}$, we can equivalently replace $B$ by $B/a^2$.

There are two other scaling invariance rules. The first is the energy scale: under the transformation $\bar t_1\rightarrow s\bar t_1,\bar t\rightarrow s \bar t,\bar t_0\rightarrow s \bar t_0, m_0\rightarrow s m_0,\mu\rightarrow s \mu$, where $\bar t_1,\bar t,\bar t_0$ can also be $t_1,t,t_0$, the result does not change. This is merely changing the unit used to measure all energies.
Another more important scaling invariant behavior related to $B$ is: under the transformation $\bar t_1\rightarrow\sqrt{s}\bar t_1,\bar t\rightarrow s \bar t$ and $B\rightarrow B/s$ or equivalently $ t_1\rightarrow\sqrt{s} t_1, t\rightarrow s t,l_{B}^2\rightarrow s l_{B}^2$, the result does not change. Although the latter form looks similar to the foregoing length rescaling by the factor $\sqrt{s}$, it is actually different, as seen from the absence of $t_0,L_z$, and can be totally independent, because we fix the length scale $a$ here. With this, one can reduce $B$ by increasing the characteristic length scale $l_0=2 t/t_1$ by a factor $\sqrt{s}$. To see the relevance of $l_0$, one can equate two typical energies $2n u_{B}=\sqrt{n}v_{B} $ in Eq.~\eqref{eq:LLs}, e.g., for the low LLs with $|n|\leq6$, whose competition in a way results in NM~I.
In this spirit, $l_0$ can be roughly taken as a relevant scale for NM~I, but \textit{not} for NM~II, because the latter does not possess an intrinsic length scale, which is exactly consistent with its persistent appearance.

The parameters $m_0,t_1,t,t_0$ for Eq.~\eqref{eq:h} are often given in the literature in units of $\mathrm{eV}$, $\mathrm{eV\text{\AA}}$, $\mathrm{eV\text{\AA}^2}$, $\mathrm{eV\text{\AA}^2}$, respectively. If we do not know the value of lattice constant $a$, we can directly use these values, set $L_z$ without specifying the number of layers, and set $l_{B}$ in real length units such as $\text{\AA}$, nm, etc., from which the real $B$ can be deduced with real values of $e,\hbar$.
On the other hand, if we are aware of $a$, we can effectively change to the $a=1$ setting: we find $\bar t_1,\bar t,\bar t_0$ first and then assume $a=1$, now we have $t_1=\bar t_1,t=\bar t,t_0=\bar t_0$ and can set $L_z=N$ by specifying the number of layers $N$, and the magnetic length $l_{B}$ is again measured in units of $a=1$, from which the real $B$ can be deduced with real values of $e,\hbar,a$. This is what we adopt in the numerical calculations. Certainly, one can also use the former way even if $a$ is known.

Below we exemplify such an estimation based on some reported realistic Weyl and Dirac semimetal materials. For the magnetic WSM EuB$_6$\cite{Nie2020}, the corresponding parameters are given as $m_0=0.16\mathrm{eV},t_0=10.4\mathrm{eV}\textrm{\AA}^2,t=31.6\textrm{eV}\textrm{\AA}^2$. We find $t_1=\hbar v_F$ from the typical Fermi velocity $v_F\sim3\times10^5\mathrm{m/s}$ and hence $\cm_0=0.11\mathrm{eV}$. Note that often $t_0/2,t/2$ are given as $M_1,M_2$ in the literature.  
For Co$_3$Sn$_2$S$_2$\cite{Muechler2020}, exploiting the scale invariance above with $a=1\text{\AA}$, another set of values is used for qualitative estimation, $m_0=1.0\mathrm{eV},t_1=1.0\mathrm{eV}\textrm{\AA},t_0=2.0\mathrm{eV}\textrm{\AA}^2,t=2.0\textrm{eV}\textrm{\AA}^2$, leading to $\cm_0=0.5\mathrm{eV}$.
A closely related Dirac semimetal Na$_3$Bi\cite{Wang2012a} gives $m_0=0.2\mathrm{eV},t_1=2.5\mathrm{eV}\textrm{\AA},t_0=21.2\textrm{eV}\textrm{\AA}^2,t=20.8\textrm{eV}\textrm{\AA}^2$. 
In these examples, we typically find $\bar t\sim 2\textrm{-}5\bar t_1$.
Therefore, the rare case of nonmonotonicity with very small $\cm_0$, given by Eq.~\eqref{eq:cond_exception}, is indeed often not realized in real materials. 
As physically expected, the typical energy scales $m_0,\bar t_1,\bar t,\bar t_0$, related to the bandwidth and so on, do not differ too much in their orders of magnitude.
Following these experimental data, from $\bar t_1=0.36\mathrm{eV},\bar t=1.54\mathrm{eV}$, consistent with the above Fermi velocity, and a typical lattice constant $a=5\text{\AA}$, we find the characteristic 
field strength $\cB_0=\hbar/el_0^2\sim30\text{T}$ for NM~I. Thus, for a significant NM~I effect, it is expected to come with an observation window roughly between $10\text{T}$ and $60\text{T}$.
Also, in these example systems, we all have $\cm_0<m_0$, and hence assures NM~I at work beside the always present NM~II. It is, however, necessary to note that this is not always guaranteed and may depend on the material choice; if $m_0<\cm_0$ is the case, NM~II will be the major contribution.

\subsection{Relation to existing experiments}\label{App:existing_experiment}
Here, we discuss the relation of our theory to the existing experiments. In the past few years, there have been various Hall effect measurements in confirmed and predicted magnetic WSMs. Among them, some show anomalously nonmonotonic magnetic field dependence, sometimes termed `nonlinear' variation as opposed to the $B$-linear OHE and the AHE due to magnetization.
They are typically observed in the quantum regime with $\omega_c\tau$ larger than or close to unity, where quantum effects appear and the major contribution is found not from extrinsic scatterings\cite{Suzuki2016,Liu2018a,Shekhar2018,Wu2023}.
These phenomena are often not well understood in the experiments. There is no clear and consistent theory or understanding, except, e.g., phenomenological fittings with both electron and hole carriers.

\begin{figure}[hbt]
\includegraphics[width=17.8cm]{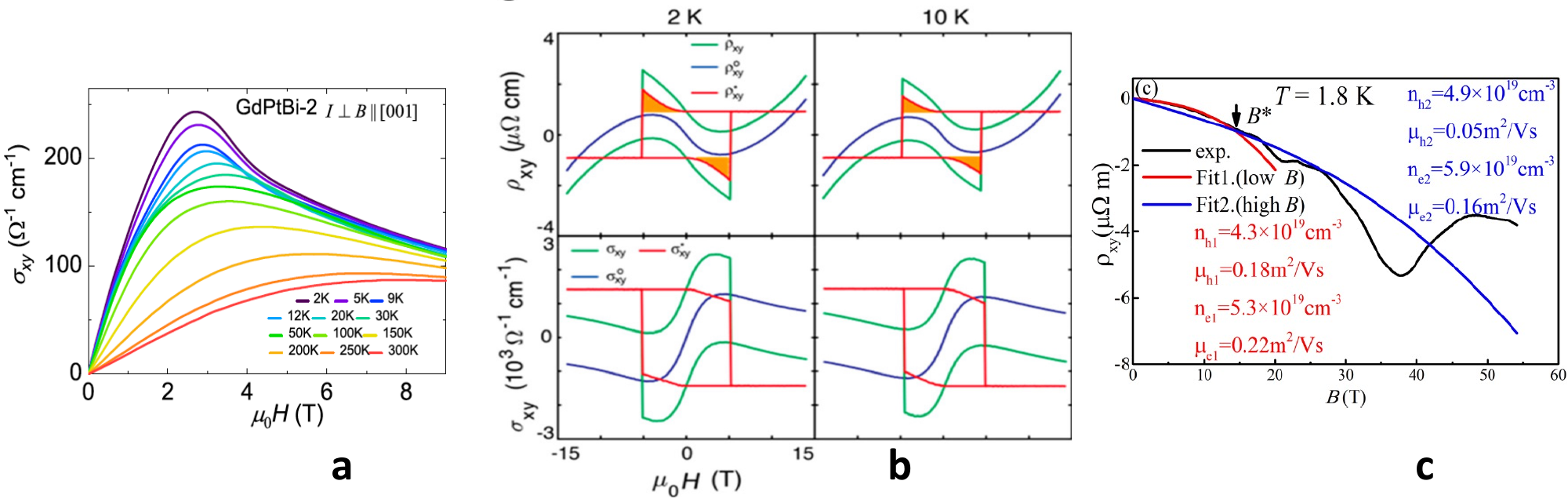}
\caption{Nonmonotonic Hall signals observed in three representative experiments. (a) Hall conductivity $\sigma_{xy}$ of GdPtBi. (b) (Top) Hall resistivity $\rho_{xy}$ and (Bottom) Hall conductivity $\sigma_{xy}$ of Co$_3$Sn$_2$S$_2$. (c) Hall resistivity $\rho_{xy}$ of PrAlSi.
(a) Adapted from Ref.~\cite{Shekhar2018} under \href{https://www.pnas.org/author-center/publication-charges/standard-pnas-license-terms}{PNAS Exclusive License to Publish}; (b) adapted from Ref.~\cite{Yang2020b} under \href{https://pubs.acs.org/page/policy/authorchoice_ccby_termsofuse.html}{Creative Commons CC BY 4.0}; (c) adapted from Ref.~\cite{Wu2023} under \href{https://pubs.acs.org/page/policy/authorchoice_ccby_termsofuse.html}{Creative Commons CC BY 4.0}.
}\label{Fig:Exp}
\end{figure}

For instance, GdPtBi\cite{Suzuki2016,Shekhar2018} and Co$_3$Sn$_2$S$_2$\cite{Yang2020b} exhibit consistently a simple nonmonotonic profile: as $B$ decreases, Hall conductivities first increase and then decrease as shown in Fig.~\ref{Fig:Exp}(a) and (b)(Bottom). This is qualitatively the same trend as, e.g., Fig.~\ref{Fig:main}(d3), which is mainly due to prominent NM~II. Refs.~\cite{Shekhar2018,Yang2020b} showing $\sigma_H$, reproduced in Fig.~\ref{Fig:Exp}(a,b), are more akin to our theoretical profile than Fig.~2(e) of Ref.~\cite{Suzuki2016} showing $\sigma_H/\sigma_{xx}$, as naturally expected. They are typically observed below $10\mathrm{T}$ or $15\mathrm{T}$, thus consistent with our prediction that NM~I will be more pronounced at higher fields. 
Furthermore, Fig.~\ref{Fig:Exp}(b) measures both $B>0$ and $B<0$ and thus includes the magnetization reversal. This WSM is considered to be due to field-induced spin-splitting with the corresponding WP separation controlled by the external field, the reversal hence means the sign reversal of $t_0,t$ of the WSM model. Accordingly, the magnetization-reversed $B<0$ region also matches our theory as per the Onsager relation Eq.~\eqref{eq:symm2}. %The profile matches the $t>0,\chi=1,\mu>0,B>0$ case. 
Two other Hall resistivity $\rho_H$ measurements in Co$_3$Sn$_2$S$_2$\cite{Liu2018a,Li2020b} are similar to Fig.~\ref{Fig:Exp}(b)(Top) and follow the expected trend. These experiments also well exemplify the $\omega_c\tau<1$ behavior at small $|B|$, where there is no divergence as mentioned in \ref{App:LLspacing} and in the main text.

On the other hand, Hall signals in PrAlSi at pulsed high fields up to $55\mathrm{T}$\cite{Wu2023} show more irregular nonmonotonic behavior with multiple nonmonotonic turns as shown in Fig.~\ref{Fig:Exp}(c). This is similar to, e.g., the curves in Fig.~\ref{Fig:main}(c3), given that nonmonotonicity is normally preserved when converted to conductivity. This is presumably related to our NM~I, which is indeed estimated to be more pronounced at relatively higher fields. Such significant nonmonotonic behavior was found to appear when $B>B^*\sim15\mathrm{T}$ in the measurement in Fig.~\ref{Fig:Exp}(c)\cite{Wu2023}, which is consistent with our estimation above.
Also resembling the theory, deviation from SdHE uniform in $1/|B|$ was seen in PrAlSi\cite{Wu2023} and also GdPtBi\cite{Suzuki2016}.
Note that other differences between the experimental data and Fig.~\ref{Fig:main}(c3) are possible for at least two reasons. i) Fig.~\ref{Fig:main}(c3) is only for $\chi=-1,b>0,\mu>0,t>0$ while measurement settings may correspond to those distinct parameters with nonmonotonicity; ii) our theory may be thought as a part of the Hall response and there remains other contribution. 

In a similar sense, although strongly hinted by the comparison, care should be taken to conclude that the anomalies in Fig.~\ref{Fig:Exp} are purely due to our predicted phenomena. While the conventional AHE, with a small relaxation time $\tau$, can be largely constant per the saturated magnetization, it is not necessarily so when $\tau$ becomes larger\cite{Nagaosa2010}. The magnetization responsible for the conventional AHE can sometimes be induced by the magnetic field, hence the field dependence of Hall signals is not always simple. 
An example is the AHE observed in EuTiO$_3$ thin films, where the nonlinear behavior during the magnetization process is attributed to the field-induced magnetization and hence band structure changes\cite{Takahashi2018}. In that scenario, the core mechanism is still the known AHE due to WP separation and Berry phase effects, although the magnetization variation affects the appearance and movement of WPs in complex ways. Therefore, in contrast to our proposal, it is more sensitive to the specific magnetism, band structure, and chemical potential in a material; also, our nonmonotonic behavior, by construction, can go well beyond the magnetization process and hence appears after the saturation, although it can also start naturally with a smaller unsaturated magnetization. These caveats aside, our prediction of the nonmonotonic Hall effect definitely provides important scenarios for understanding existing experiments and guiding future ones.

\subsection{Chirality selectiveness with multiple WP pairs}\label{App:selective}

\begin{figure}[hbt]
\includegraphics[width=8.8cm]{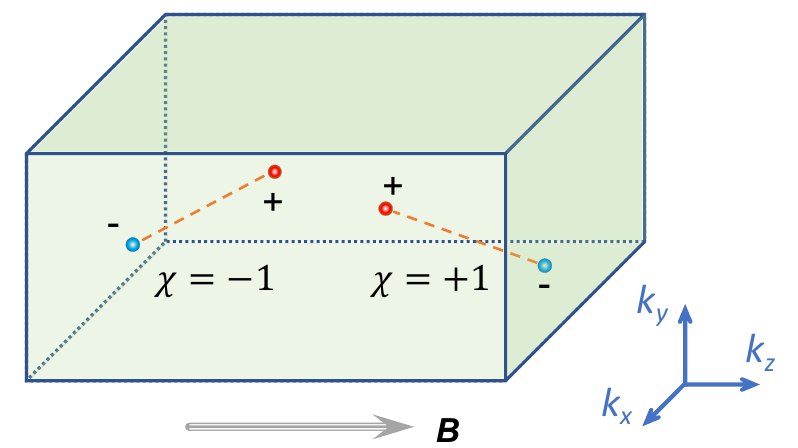}
\caption{Schematic representation of a case with more than one pair of WPs in the 3D momentum space. Positive-charge WPs are in red while negative-charge WPs are in blue. The chirality parameter $\chi$ is indicated for the two WP pairs. 
}\label{Fig:schematic}
\end{figure}

As mentioned in the main text, magnetic WSM materials may have more than one pair of WPs, although the minimal number allowed is one. Consider Fig.~\ref{Fig:schematic} with two pairs as a representative example. Note that, in order to contribute to the nonmonotonic Hall effect, the magnetic field $\bB$ does not need to be strictly parallel to the WP alignment in a pair, because it is the momentum slices normal to $\bB$ that count. The left WP pair represents a spatially tilted version of our original one-pair Hamiltonian $H(\chi=-1)$ in the main text or Eq.~\eqref{eq:H}, which is similar to the configuration in Fig.~\ref{Fig:main}(a). Once the chirality parameter $\chi$ in $H$ or its continuum counterpart Eq.~\eqref{eq:h_main} is reversed, both charges of the two WPs will be reversed since $\chi$ controls an odd number of the signs of Fermi velocity components. Therefore, we have the right pair as indicated in Fig.~\ref{Fig:schematic}.

For the well-known AHE of magnetic WSM due to the Berry phase effect, e.g., $\sigma_{xy}$ when $\bB=\bzero$, the transport contribution from the left and right pairs will cancel out to a considerable extent, because their Berry phase fluxes are in gross opposite. However, this is not what would happen for the nonmonotonic Hall effect we proposed. Consider the general condition Eq.~\eqref{eq:NMLL_0_condition} for nonmonotonic behavior, only one pair in the two is at work for a given chemical potential. For instance, when $\mu>0$, the $\chi=+1$ pair does not contribute to nonmonotonicity for the configuration shown in Fig.~\ref{Fig:schematic}. 

This feature, therefore, constitutes an interesting selectiveness of the chirality $\chi$: only those WP pairs compatible with Eq.~\eqref{eq:NMLL_0_condition} are able to contribute to the nonmonotonicity. Firstly, this beneficially avoids any cancellation as in the AHE. Secondly, this might help determine the configuration of WP pairs in experiments. For instance, if one detects a clear nonmonotonic Hall effect with some given magnetic field and chemical potential, e.g., $\bB=B\hat{z},B>0,\mu>0$, it is valid to presume the existence of at least one WP pair in a configuration similar to the left one in Fig.~\ref{Fig:schematic}. Once we reverse the direction of $\bB$, the absence (presence) of the nonmonotonic Hall effect would exclude (confirm) the existence of some WP pair in a configuration similar to the right one in Fig.~\ref{Fig:schematic}. If the sign of the chemical potential could further be changed by certain experimental means, this feature for finding pair configurations would become even more powerful and accurate.

\end{document}